\RequirePackage{ifpdf}
\ifpdf 
\documentclass[pdftex]{sigma}
\else
\documentclass{sigma}
\fi

\numberwithin{equation}{section}

\begin{document}

\allowdisplaybreaks

\renewcommand{\thefootnote}{$\star$}

\renewcommand{\PaperNumber}{055}

\FirstPageHeading

\ShortArticleName{Space-Time Dif\/feomorphisms in Noncommutative Gauge Theories}

\ArticleName{Space-Time Dif\/feomorphisms\\ in Noncommutative Gauge Theories\footnote{This paper is a
contribution to the Special Issue on Deformation Quantization. The
full collection is available at
\href{http://www.emis.de/journals/SIGMA/Deformation_Quantization.html}{http://www.emis.de/journals/SIGMA/Deformation\_{}Quantization.html}}}

\Author{Marcos ROSENBAUM, J. David VERGARA and L. Rom\'an JUAREZ}

\AuthorNameForHeading{M. Rosenbaum, J.D. Vergara and L.R. Juarez}

\Address{Instituto de Ciencias Nucleares,
Universidad Nacional Aut\'onoma de M\'exico, \\
 A. Postal 70-543, M\'exico D.F., M\'exico}
\Email{\href{mailto:mrosen@nucleares.unam.mx}{mrosen@nucleares.unam.mx}, \href{mailto:vergara@nucleares.unam.mx}{vergara@nucleares.unam.mx},\\
\hspace*{14mm}\href{mailto:roman.juarez@nucleares.unam.mx}{roman.juarez@nucleares.unam.mx}} %

\ArticleDates{Received April 11, 2008, in f\/inal form June 25,
2008; Published online July 16, 2008}

\Abstract{In previous work 
[Rosenbaum M. et al.,
   {\it J.\ Phys.\ A: Math. Theor.}  {\bf 40} (2007), 10367--10382]
we have shown how for canonical parametrized f\/ield theories,
where space-time is placed on the same footing as the other f\/ields in the theory,
the  representation of space-time dif\/feomorphisms  provides a very convenient scheme for
analyzing the induced twisted deformation of these dif\/feomorphisms, as a result of the space-time noncommutati\-vity. However,
 for gauge f\/ield theories (and of course also for canonical geometrodynamics) where the Poisson brackets of the constraints
explicitely depend on the embedding variables, this Poisson algebra cannot be connected directly with a representation of the
complete Lie algebra of space-time dif\/feomorphisms, because not all the f\/ield variables turn out to have a~dynamical character
[Isham C.J.,  Kucha\v{r} K.V., {\it Ann. Physics} {\bf 164} (1985), 288--315, 316--333].
Nonetheless, such an homomorphic mapping can be recuperated by
f\/irst modifying the original action  and then adding additional constraints in the formalism in order to retrieve the original theory, as shown by Kucha\v {r} and
Stone for the case of the parametrized Maxwell f\/ield in
[Kucha\v{r} K.V., Stone S.L., {\it Classical Quantum Gravity}  {\bf 4} (1987), 319--328].
Making use of a combination of all of these ideas, we are therefore able to apply our canonical reparametrization approach in order to
 derive the deformed Lie algebra
of the noncommutative space-time dif\/feomorphisms as well as to consider how gauge transformations act on the twisted
algebras of gauge and particle f\/ields. Thus, hopefully, adding clarif\/ication on some
outstanding issues in the literature concerning the symmetries for gauge theories in noncommutative space-times.}


\Classification{70S10; 70S05; 81T75}

\renewcommand{\thefootnote}{\arabic{footnote}}
\setcounter{footnote}{0}

\section{Introduction}

\looseness=1
Within the context of quantum f\/ield theory, a considerable amount of work has been done
recently dealing with quantum f\/ield theories in noncommutative space-times (NCQFT). One
of the most relevant issues in this area is related to the symmetries
under which these noncommutative systems are invariant. The most recent
contention being that NCQFT are invariant under global ``twisted symmetries"
(see, e.g., \cite{Chaichian1}). This criterion has been extended
to the case of the
twisting  of local symmetries, such as dif\/feomorphisms
\cite{Wess1}, and this has been used to propose some noncommutative
theories of gravity \cite{Wess1, Wess2,Wess2+}. Another possible
extension of this idea is to consider the construction of
noncommutative gauges theories with an arbitrary gauge group
\cite{vass, Wess3}. Regarding this latter line of research there is, however,
some level of controversy as to whether it is possible to construct
twisted gauge symmetries
\cite{Chaichian2, Chaichian3, Banerjee}. In this
work we address this issue from the point of view of canonically
reparametrized f\/ield theories. It is known indeed that for the
case of f\/ield theories with no internal symmetries, it is pos\-sib\-le
to establish, within the framework of the canonical parametrization,
an  anti-homorphism between the Poisson algebra of the constraints on
the phase space of the system and the algebra of space-time dif\/feomorphisms \cite{isham1,isham2}.
Using this anti-homomorphism we were able in~\cite{nos1} to show
how the deformations of the algebra of constraints, resulting from
space-time noncommutativity at the level of the quantum mechanical
mini-superspace, are ref\/lected on the twisting of the algebra of the f\/ields
as well as in the Lie algebra of the twisted dif\/feomorphisms  and in the
ensuing \mbox{twisting} of the original symmetry group of the theory.
However, as it has also been noted by Isham and Kucha\v {r} in \cite{isham1,isham2},
for the case of gauge theories, there are some dif\/f\/iculties in representing space-time dif\/feomorphisms
by an anti-homomorphic mapping into the Poisson algebra of the
dynamical variables on the extended phase space of the canonically
reparametrized theory, due to the fact that because of the additional internal symmetries
some components of the f\/ield loose their dynamical character and appear as Lagrange multipliers
in the for\-malism.

Nonetheless, as it was exemplif\/ied in \cite{Kuchar2}
for the case of the  parametrized Maxwell f\/ield, such dif\/f\/iculties can
be circumvented and the desired mapping made possible by adding some terms
to the original action and some additional constraints in order to recover
the original features of the theory.

\looseness=1
Making therefore use of the specif\/ic results derived by Kucha\v {r} and Stone in \cite{Kuchar2}
for the parametrized Maxwell f\/ield and the re-established mapping between
the space-time dif\/feo\-mor\-phisms and the Poisson algebra of the modif\/ied theory,
together with our previous results in~\cite{ros1}~-- whereby
noncommutativity in f\/ield theory, manifested as the twisting of the algebra of f\/ields, has a dynamical origin
in the quantum mechanical mini-superspace which, for f\/lat Minkowski space-time, is related to an extended Weyl--Heisenberg group --
and including this results
into a generalized symplectic structure of the parametrized f\/ield theory \cite{nos1}, we show here how our approach can be extended to gauge f\/ield theories
thus allowing us to derive the deformed Lie algebra
of the noncommutative space-time dif\/feomorphisms, as well as to consider how the gauge transformations act on the twisted
algebras of gauge and particle f\/ields. Hopefully this approach will help shed some additional univocal
light on the above mentioned controversy.

The paper is organized as follows: In Section~\ref{sec2} we review the essential aspects of the construction
of canonical parametrized f\/ield theories and representations of space-time dif\/feomorphisms,
following \cite{isham1, isham2, Kuchar, Kuchar1}. In Section~\ref{sec3} we show how the formalism can be extended to the
case of parametrized gauge f\/ield theories by
making use of the ideas formulated in \cite{Kuchar2} in the context of Maxwell's
electrodynamics. Section~\ref{sec4} summarizes in the language of Principal Fiber Bundles (PFB) some of the
basic aspects of the theory of gauge transformations which will be needed in the later part of the
work. In Section~\ref{sec5} we combine the results of the previous sections in order to extend the formalism
to the noncommutative space-time
case, by deforming the symplectic structure of the theory to account for the noncommutativity
of the space-time embedding coordinates.

We thus derive a deformed algebra of constraints in terms of
Dirac-brackets which functionally satisfy the same Dirac relations
as those for the commutative case and can therefore be related
anti-homomorphically to a Lie algebra of generators of twisted
space-time dif\/feomorphisms. On the basis of  these results we further
show how, in order to preserve the consistency of the algebra of
constraints, the Lie algebra of these generators of space-time
dif\/feomorphisms and those of the gauge symmetry are in turn related.

Finally by extending the algebra of twisted dif\/feomorphisms to its
universal covering, it was given an additional Hopf structure which allowed
us to relate the twisting of symmetry of the theory to the Drinfeld twist.

\section[Space-time diffeomorphisms in parametrized gauge theories]{Space-time dif\/feomorphisms in parametrized gauge theories}\label{sec2}

As it is well known, see e.g.\ \cite{isham1, isham2}, for Poincar\'e invariant f\/ield theory
on a f\/lat Minkowskian background, each generator of the Poincar\'e Lie algebra, represented by a dynamical
variable on the phase-space of the f\/ield, is mapped homomorphically into the Poisson bracket algebra of these
dynamical variables.

On a curved space-time background f\/ield theories are not Poincar\'e invariant but, by a pa\-ra\-met\-ri\-za\-tion
consisting of extending the phase-space by adjoining to it the embedding
variables, they can be made invariant under arbitrary
space-time dif\/feomorphisms \cite{dirac, Kuchar3}. Hence space-time parameters are raised to the level of f\/ields
on the same footing as the original f\/ields in the theory. Moreover, in this case it can also be shown \cite{isham1} that:

a) An anti-homomorphic mapping can be established  from the Poisson
algebra of dynamical variables on the extended phase-space and the
Lie algebra $\pounds\,\text{dif\/f}\,\mathcal M$ of arbitrary
space-time dif\/feomorphisms. Thus, \begin{gather*}
\{  H_\tau[\xi],
H_\tau[\eta] \} = - H_\tau[\pounds_\xi \eta], \end{gather*} where $\xi,\eta
\in \text{\pounds\, dif\/f}\,\mathcal M$ are two complete space-time
Hamiltonian vector f\/ields on $\mathcal M$, $H_\tau[\xi]:=\int_\Sigma
d\boldsymbol\sigma \;\xi^\alpha {\mathcal H}_\alpha$, and $
{\mathcal H}_\alpha$ are the constraints (supermomenta and
superHamiltonian) of the theory, satisfying the Dirac vanishing
Poisson bracket algebra \begin{gather}\label{dialg} \{{\mathcal
H}_{\alpha}(\boldsymbol\sigma) , {\mathcal
H}_{\beta}(\boldsymbol\sigma')\}\simeq 0. \end{gather}

b) The Poisson brackets of the canonical variables representing the $\pounds\,\text{dif\/f} \,\mathcal M$ correctly
induce the displacements of embeddings accompanied by the evolution of the f\/ield variables, predicted by the f\/ield equations.

For the prescribed pseudo-Riemannian background $\mathcal M$,
equipped with coordinates $X^\alpha$, re\-pa\-ra\-met\-ri\-za\-tion involves a
foliation $\Sigma \times {\mathbb R}$ of this space-time, where ${\mathbb R}$
is a temporal direction labeled by a parameter $\tau$ and $\Sigma$
is a space-like hypersurface of constant $\tau$, equipped with
coordinates~$\sigma^a$ $(a=1,2,3)$, and embedded in the space-time
4-manifold by means of the mapping \begin{gather*}
X^\alpha =
X^{\alpha}(\sigma^a). \end{gather*}
This hypersurface is assumed to be spacelike with respect to the metric $g_{\alpha\beta}$ on $\mathcal M$, with signature $(-,+,+,+)$.

Let now the embedding functionals ${X^\alpha}_a (\boldsymbol\sigma, X):= \frac{\partial X^\alpha(\boldsymbol\sigma)}{\partial\sigma^a}$ and $n^\alpha(\boldsymbol\sigma, X)$,
 def\/ined by
\begin{gather}\label{basis}
g_{\alpha\beta}{X^\alpha}_a n^\beta =0,\qquad \text{and}\qquad g_{\alpha\beta}n^\alpha n^\beta =-1,
\end{gather}
be an anholonomic basis consisting of tangent vectors to the hypersurface and unit normal, respectively.

We can therefore write the constraints ${\mathcal H}_\alpha$ as
\begin{gather*}
{\mathcal H}_\alpha = -{\mathcal H}_\perp n_\alpha + {\mathcal H}_a {X_\alpha}^a,
\end{gather*}
where ${\mathcal H}_\perp$ and ${\mathcal H}_a$ are the super-Hamiltonian and super-momenta constraints, respectively.
\mbox{Using} this decomposition the Dirac relations (\ref{dialg}) can be written equivalently as
\begin{gather*}
\{{\mathcal H}_{\perp} (\boldsymbol\sigma) , {\mathcal H}_{\perp}
({\boldsymbol\sigma}')\} = \sum_{a=1}^{3} \gamma^{ab}{\mathcal H}_{b} (\boldsymbol\sigma)
 \partial_{\sigma^a}\delta(\boldsymbol\sigma-{\boldsymbol\sigma}')-(\boldsymbol\sigma\leftrightarrow{\boldsymbol\sigma}'),\nonumber\\
\{{\mathcal H}_{a} (\boldsymbol\sigma), {\mathcal H}_b
({\boldsymbol\sigma}')\} = {\mathcal H}_b
(\boldsymbol\sigma)\partial_{\sigma^a}
\delta(\boldsymbol\sigma-{\boldsymbol\sigma}') +{\mathcal H}_a
({\boldsymbol\sigma}')\partial_{\sigma^b}
\delta(\boldsymbol\sigma-{\boldsymbol\sigma}'),
\\
\{{\mathcal H}_{a} (\boldsymbol\sigma), {\mathcal H}_\perp
({\boldsymbol\sigma}')\} = {\mathcal H}_{\perp}
(\boldsymbol\sigma)\partial_{\sigma^a}
\delta(\boldsymbol\sigma-{\boldsymbol\sigma}'),\nonumber
\end{gather*}
where $\gamma ^{ab}$ is the inverse of the spatial metric
\begin{gather*}
\gamma_{ab}(\boldsymbol\sigma,X):= g_{\alpha\beta}(X(\boldsymbol\sigma)) {X^\alpha}_a {X^\beta}_b .
\end{gather*}

Also, as a consequence of the antihomomorphism between the Poisson algebra of the constraints and
$\text{\pounds\, dif\/f} \,{\mathcal M}$ we can write
\begin{gather*}
H_\tau[\xi] \leadsto\hat H_{\tau}[\xi]\equiv
\delta_{\xi}=\xi^\alpha(X(\tau,\boldsymbol\sigma))
\left.\frac{\partial}{\partial
X^\alpha}\right|_{X(\tau,\boldsymbol\sigma)}.
\end{gather*}
Indeed, since $[\eta ,\rho ]=\pounds_\eta \rho $ we have
\begin{gather*}
[\delta_\eta , \delta_\rho ]\phi =  \delta_{\pounds_\eta \rho} \phi=\hat H_{\tau} [\pounds_{\eta} \rho]\triangleright \phi
\cong  \{\phi, H_{\tau} [\pounds_{\eta} \rho]\} \nonumber\\
\phantom{[\delta_\eta , \delta_\rho ]\phi}{} = \hat H_\tau [\eta] \triangleright[\hat H_\tau [\rho] \triangleright\phi]-
\hat H_\tau [\rho] \triangleright[\hat H_\tau [\eta] \triangleright\phi]  \nonumber  \\
\phantom{[\delta_\eta , \delta_\rho ]\phi}{}  \cong \{\{\phi,  H_\tau [\rho]\}, H_\tau [\eta]\}-\{\{\phi, H_\tau [\eta]\} , H_\tau[\rho]\}
=-\{\phi, \{  H_\tau[\eta], H_\tau[\rho]\}\}
\end{gather*}
after resorting to the Jacobi identity and where $\phi$ is some f\/ield function in the theory.

Making use of this antihomomorphism as well as of the dynamical
origin of $\star$-noncommutati\-vi\-ty in f\/ield theory from quantum
mechanics exhibited in \cite{ros1}, we have considered in
\cite{nos1} the extension of the reparametrization formalism and the
canonical representation of space-time dif\/feomorphisms to the study
of f\/ield theories on noncommutative space-times. More specif\/ically,
in that paper we discussed the particular case of a Poincar\'e
invariant scalar f\/ield immersed on a f\/lat Minkowskian background,
and showed that the deformation of the algebra of constraints due to
the incorporation of a symplectic structure in the theory originated
the Drinfeld twisting of that isometry. However, although the
formalism developed there can be extended straightforwardly to any
f\/ield theory with no internal symmetries, for the case of
parametrized gauge theories some additional complications arise, as
pointed out in \cite{isham2} and \cite{Kuchar2}, due to the fact
that the components of the gauge f\/ield perpendicular to the
embedding are not dynamical but play instead the role of Lagrange
multipliers which are not elements of the extended phase space and
therefore can not  be turned into dynamical variables by canonical
transformations. To do so, and recover the anti-homomorphism between
the algebra of space-time dif\/feomorphisms and the Poisson algebra of
constraints it is necessary to impose additional Gaussian
conditions. The simplest case where such a procedure can be
exhibited is the parametrized electromagnetic f\/ield. This has been
very clearly elaborated in \cite{Kuchar2}, so we shall only review
those aspects of that work needed for our presentation.

\section[Parametrized Maxwell field and canonical representation of space-time diffeomorphisms]{Parametrized Maxwell f\/ield and canonical representation\\ of space-time dif\/feomorphisms}\label{sec3}

Consider a source-free Maxwell f\/ield in a prescribed
pseudo-Riemannian space-time represented by the action
\begin{gather}\label{SM}
    S=-\frac{1}{4} \int d^4X \sqrt{-g}\: g^{\mu\nu} g^{\alpha\beta}
    F_{\mu\alpha}F_{\nu\beta},
\end{gather}
where $F_{\mu\alpha}=A_{[\mu,\alpha]}:=A_{\mu,\alpha}
-A_{\alpha,\mu}$. In the canonical treatment of the evolution of a
f\/ield one assumes it to be def\/ined on a space-like 3-hypersurface
$\Sigma$, equipped with coordinates $ \boldsymbol\sigma$, which is
embedded in the space-time manifold $\mathcal M$ by the mapping
\begin{gather*}
    X^\mu :(\boldsymbol\sigma)=X^\mu(\sigma^a),\qquad a=1,2,3.
\end{gather*}
By adjoining the embedding variables to the phase space of the f\/ield
results in a parametrized f\/ield theory where the space-time
coordinates have been promoted to the rank of f\/ields. In terms of
the space-time coordinates $\sigma^\alpha=(\tau,\boldsymbol\sigma)$
determined by the foliation $\mathcal{M}= \mathbb{R}\times \Sigma$,
the action~(\ref{SM}) becomes
\begin{gather}\label{SM1}
    S=-\frac{1}{4} \int d\tau d^3\sigma \  \sqrt{-\bar g} \bar g^{\mu\nu}\bar g^{\alpha\beta}
    F_{\mu\alpha}F_{\nu\beta},\qquad t\in \mathbb{R},
\end{gather}
with the inverse metric  $\bar g^{\alpha\beta}$ given by
\begin{gather*}
\bar g^{\alpha\beta}= \frac{\partial\sigma^\alpha}{\partial X^\mu}
\frac{\partial \sigma^\beta}{\partial X_\mu},
\end{gather*}
which can be therefore seen as a function of the coordinate f\/ields.
In (\ref{SM1}) $\bar g:=\det(\bar g_{\mu\nu})$ where $\sqrt{-\bar
g}=J$
is the Jacobian of the transformation.

In order to carry out the Hamiltonian analysis of the action
(\ref{SM1}), we def\/ine in similar way to (\ref{basis}) the tangent
vectors to $\Sigma$, $X^\alpha _{\;a}$ and the unit normal $n^\alpha
= -(-\bar g^{00})^{-\frac{1}{2}}\bar g^{0\rho}\frac{\partial
X^\alpha}{\partial\sigma^\rho}$. We thus arrive at
\begin{gather}\label{Nocom1}
    S[X^\mu,P_\mu,A_a,\pi^a, A_\perp]=\int d\tau d^3\sigma \big(P_\alpha \dot X^\alpha + \pi^a \dot A_a -N\Phi_0 -N^a \Phi_a -M
    G\big),
\end{gather}
where $N$ and $N^a$ are the lapse and shift components of the
deformation vector $N^\alpha:=\partial X^\alpha / \partial\tau $, $
M=NA_\perp -N^a A_a$, and $A_a :=X^\alpha_{\;a}A_\alpha$,
$A_\perp := -n^\beta A_\beta $ are the tangent and normal
projections of the gauge potential. The constraints $\Phi_0$, $\Phi_a$
and $G$ in (\ref{Nocom1}) are def\/ined by:
\begin{gather}
\Phi_0 = P_\alpha n^\alpha + \frac{1}{2} \gamma^{-1/2}
\gamma_{ab}\pi^a \pi^b + \frac{1}{4}
\gamma^{1/2}\gamma^{ac}\gamma^{bd}F_{ab} F_{cd},\nonumber\\
  \Phi_a      = P_\alpha X^\alpha_{\ ,a} +F_{ab}\pi^b,\qquad
  G = \pi^a_{\ ,a},\label{Cons1}
\end{gather}
where $\gamma_{ab}$, $\gamma$ are the metric components on $\Sigma$ and their determinant, respectively.
These constraints satisfy the relations:
\begin{gather}
\{\Phi_{0}(\boldsymbol\sigma) + A_\perp(\boldsymbol\sigma)G(\boldsymbol\sigma),
 \Phi_{0}(\boldsymbol\sigma') + A_\perp(\boldsymbol\sigma')G(\boldsymbol\sigma')\}
=\big[\gamma^{ab}(\boldsymbol\sigma)\Phi_b(\boldsymbol\sigma)
 \!+\!\gamma^{ab}(\boldsymbol\sigma')\Phi_b(\boldsymbol\sigma') \big]
\delta_{,a}(\boldsymbol\sigma{,}\boldsymbol\sigma'),\nonumber\\
\{\Phi_{a}(\boldsymbol\sigma) - A_a(\boldsymbol\sigma) G(\boldsymbol\sigma),
 \Phi_{b}(\boldsymbol\sigma') -
A_b(\boldsymbol\sigma') G(\boldsymbol\sigma')\}\nonumber\\
\qquad{}=(\Phi_{b}(\boldsymbol\sigma) - A_b(\boldsymbol\sigma)G(\boldsymbol\sigma))\delta_{,a}(\boldsymbol\sigma,\boldsymbol\sigma')
+\left(\Phi_a(\boldsymbol\sigma') -A_a(\boldsymbol\sigma')G(\boldsymbol\sigma')\right)
\delta_{,b}(\boldsymbol\sigma,\boldsymbol\sigma'),\nonumber\\
\{\Phi_{a}(\boldsymbol\sigma) - A_a(\boldsymbol\sigma)G(\boldsymbol\sigma),
 \Phi_{0}(\boldsymbol\sigma') + A_{\perp}(\boldsymbol\sigma')G(\boldsymbol\sigma')\}=\left(\Phi_0(\boldsymbol\sigma) +A_{\perp}(\boldsymbol\sigma)G(\boldsymbol\sigma)\right)\delta_{,a}(\boldsymbol\sigma,\boldsymbol\sigma'),
\nonumber\\
\{\Phi_{0}(\boldsymbol\sigma), G(\boldsymbol\sigma')\}= 0,\qquad \{\Phi_{a}(\boldsymbol\sigma), G(\boldsymbol\sigma')\}= 0.
\label{algebra1.1}
\end{gather}
 From here we see that the Gauss constraint $G$
is needed to achieve the closure of the algebra of the super-Hamiltonian
and super-momenta constraints, $\Phi_0$, $\Phi_a$,
under the Poisson-brackets. However, because of the gauge
invariance implied by the Gauss constraint $G\approx 0$, the scalar
potential $A_\perp $ occurs in~(\ref{algebra1.1}) not as a dynamical
variable but as a Lagrange multiplier. The end result of this mixing
of constraints and consequent foliation dependence of the space-time
action in gauge theories, is that the super-Hamiltonian,
\begin{gather*}
n^\alpha {\mathcal H}_{\alpha}={\mathcal
H}_{\perp}:=\Phi_{0}(\boldsymbol\sigma) +
A_\perp(\boldsymbol\sigma)G(\boldsymbol\sigma),
\end{gather*}
and the supermomenta,
\begin{gather*}
X^{\alpha}_a {\mathcal H}_{\alpha}={\mathcal H}_{a}
:=\Phi_{a}(\boldsymbol\sigma) -
A_{a}(\boldsymbol\sigma)G(\boldsymbol\sigma),
\end{gather*}
constraints do not satisfy the Dirac closure relations (\ref{dialg})
$( \{{\mathcal H}_{\alpha}(\boldsymbol\sigma),{\mathcal
H}_{\beta}(\boldsymbol\sigma')\}\simeq0 )$, so we do not have a
direct homomorphic map from the Poisson brackets algebra of
constraints into the Lie algebra of space-time dif\/feomorphisms for
such theories. Nonetheless, this dif\/f\/iculty can be circumvented by
turning the scalar potential into a canonical momentum $\pi$ (via
the relation $\pi= \sqrt{\gamma} A_\perp$) conjugate to a
supplementary scalar f\/ield $\psi$ and prescribing their dynamics by
imposing the Lorentz gauge condition. The new super-Hamiltonian and
super-momenta
\begin{gather}
  ^\ast {\mathcal H}_{\perp}:= {\mathcal H}_{\perp} - \sqrt{\gamma} \gamma ^{ab}\psi_{,a} A_{b} ,\qquad
^\ast{\mathcal H}_a : = {\mathcal H}_a + \pi \psi_{,a} ,\label{modsupham}
\end{gather}
of the modif\/ied theory satisfy the Dirac closure relations, and the
mapping $\xi\to  {^\ast} H_\tau [\xi]= \int_{\Sigma}
d\boldsymbol\sigma'\ \xi^\alpha(X(\boldsymbol\sigma'))\
^\ast{\mathcal H}_\alpha$ results in the desired anti-homomorphism:
\begin{gather}\label{antihomo} \{ {^\ast} H_\tau[\xi], {^\ast} H_\tau[\rho]
\}=-{^\ast} H_\tau[\pounds_{\xi}\rho], \end{gather} from the Lie algebra
$\pounds \,\text{dif\/f}\, {\mathcal M}\ni \xi, \rho $ into the Poisson
algebra of the constraints on the extended phase space $A_a$,
$\pi^a$, $\psi$, $\pi$, $X^\alpha$, $P_\alpha$ of the modif\/ied
electrodynamics with the space-time action: \begin{gather}\label{modelect}
S(\phi, \psi) = \int_{\mathcal M}d^4 X \sqrt {-g}
\left(-\frac{1}{4}F^{\alpha\beta} F_{\alpha\beta} +
\psi_{,\alpha}g^{\alpha\beta} A_\beta \right). \end{gather} Note however that
in order to recover Maxwell's electrodynamics from the dynamically
minimal modif\/ied action (\ref{modelect}), one needs to impose the
additional primary and secondary constraints \begin{gather}\label{gcons7}
C(\boldsymbol\sigma):=\psi(\boldsymbol\sigma)\approx 0, \qquad
G(\boldsymbol\sigma)\approx 0 \end{gather}
 on the phase space data. In this way, the new algebra of constraints
leading to vacuum electrodynamics from (\ref{modelect}) is:
\begin{gather}
\{ ^\ast{{\mathcal H}}_{\perp}(\boldsymbol\sigma) , ^\ast{{\mathcal H}}_{\perp}(\boldsymbol\sigma')\}
=\gamma^{ab}(\boldsymbol\sigma)\; {^\ast{{\mathcal H}}_{b}}(\boldsymbol\sigma)\delta_{,a}(\boldsymbol\sigma,\boldsymbol\sigma')
-(\boldsymbol\sigma\leftrightarrow \boldsymbol\sigma'),\nonumber\\
\{ ^\ast{ {\mathcal H}}_{a}(\boldsymbol\sigma) , ^\ast{{\mathcal H}}_{\perp}(\boldsymbol\sigma')\}
= {^\ast{{\mathcal H}}_{\perp}}(\boldsymbol\sigma)\delta_{,a}(\boldsymbol\sigma,\boldsymbol\sigma'),\nonumber\\
\{ ^\ast{{\mathcal H}}_{a}(\boldsymbol\sigma) , ^\ast{{\mathcal H}}_{b}(\boldsymbol\sigma')\}
= {^\ast{{\mathcal H}}_{b}}(\boldsymbol\sigma)\delta_{,a}(\boldsymbol\sigma,\boldsymbol\sigma')
-(a\boldsymbol\sigma\leftrightarrow b\boldsymbol\sigma'),\nonumber\\
\{ C(\boldsymbol\sigma) , ^\ast{ {\mathcal H}}_{\perp}(\boldsymbol\sigma')\}
= (\gamma)^{-\frac{1}{2}}(\boldsymbol\sigma) G(\boldsymbol\sigma)\delta(\boldsymbol\sigma,\boldsymbol\sigma'),\nonumber\\
\{ C(\boldsymbol\sigma) , ^\ast{{\mathcal H}}_{a}(\boldsymbol\sigma')\}
=  C_{,a}(\boldsymbol\sigma)\delta(\boldsymbol\sigma,\boldsymbol\sigma'),\nonumber\\
\{ G(\boldsymbol\sigma) , ^\ast{ {\mathcal H}}_{\perp}(\boldsymbol\sigma')\}
= \left((\gamma)^{\frac{1}{2}}(\boldsymbol\sigma) \gamma^{ab}(\boldsymbol\sigma)
C_{,b}(\boldsymbol\sigma)\delta(\boldsymbol\sigma,\boldsymbol\sigma')\right)_{,a},\nonumber\\
\{ G(\boldsymbol\sigma) , ^\ast{{\mathcal H}}_{a}(\boldsymbol\sigma')\}
=  \left(G(\boldsymbol\sigma)\delta(\boldsymbol\sigma,\boldsymbol\sigma')\right)_{,a}.\label{diracb2}
\end{gather}

This Poisson algebra implies that once the constraints
(\ref{gcons7}) are imposed on the initial data they are preserved in
the dynamical evolution generated by the total Hamiltonian
associated with (\ref{modelect}), so that if the derivations $
^\ast\hat H_\tau[\xi]:=\delta_\xi $ representing space-time
dif\/feomorphisms start evolving a point of the extended phase space
lying on the intersection of the constraint surfaces
\begin{gather*}
^\ast{\mathcal H}_{\perp}
(\boldsymbol\sigma)\approx 0 \approx ^\ast{\mathcal H}_a
(\boldsymbol\sigma)\qquad \text {and}\qquad
C(\boldsymbol\sigma):=\psi(\boldsymbol\sigma)\approx 0\approx
G(\boldsymbol\sigma), \end{gather*}
the point will keep moving along this
intersection.

In summary, we have seen that for canonically parametrized f\/ield
theories with gauge symmetries in addition to space-time symmetries
the Poisson algebra of the constraints does not agree with the Dirac
relations and, therefore, cannot be directly interpreted as
representing  the Lie algebra  of the generators of space-time
dif\/feomorphisms. The reason being that because of the gauge
invariance there are additional constraints in the theory which
cause that not all the relevant variables are canonical variables.
Following the arguments in \cite{Kuchar2} for the case of the
electromagnetic f\/ield, we have seen that these dif\/f\/iculties can be
circumvented by complementing the original action (\ref{SM}) with
the addition of a term, containing the scalar f\/ield $\psi$, that
enforces the Lorentz condition, so the modif\/ied action is given by
(\ref{modelect}). Varying this action with respect to the gauge
potential $A_\alpha$ gives \begin{gather}\label{max1} \frac{1}{2}(|{\sqrt
g}|^{\frac{1}{2}} F^{\alpha\beta})_{,\beta}= |{\sqrt
g}|^{\frac{1}{2}} g^{\alpha\beta}\psi_{,\beta} , \end{gather} which
therefore implies that the modif\/ied action introduces a source term
into the Maxwell equations, so the dynamical theory resulting from
(\ref{modelect}) is not the same as Maxwell's electrodynamics in
vacuum. It is interesting to observe, parenthetically, that the
charge source on the right of (\ref{max1}) is a real f\/ield and not a
complex one as one would have expected. The dynamical character of
$\psi$, however, is evident when dif\/ferentiating this last equation
with respect to $X^\alpha$ whereby, due to the vanishing of the left
side, this f\/ield must satisfy the wave equation \begin{gather*}
{\psi_{,\alpha}}^{,\alpha}=0. \end{gather*} Consequently, in order to recover
Maxwell's electrodynamics it was required that $\psi$ vanish or at
least that it is a space-time constant. This was achieved by simply
imposing additional constraints on the phase space data, given by
(\ref{gcons7}), which (c.f.\ equation~(\ref{gauss4}) in the next
section) implies loosing the generator of gauge transformations.
This procedure, and its generalization to the case of non-Abelian
Yang--Mills f\/ields then allows (still within the canonical group
theoretical framework) to undo the projection and replace the
Poisson bracket relations
(\ref{algebra1.1}) by the genuine Lie algebra  $\pounds\,\text{dif\/f}\, \mathcal M $ of
space-time dif\/feomorphisms.

Note that even though the algebra in  (\ref{diracb2}) involves derivatives of the constraints
$G(\boldsymbol\sigma)$ and~$C(\boldsymbol\sigma)$, these derivatives can
be removed by simply using the identity
\[
J(\boldsymbol\sigma')\delta_{,a}(\boldsymbol\sigma, \boldsymbol\sigma')= J(\boldsymbol\sigma)
\delta_{,a}(\boldsymbol\sigma, \boldsymbol\sigma')
+J_{,a}(\boldsymbol\sigma)\delta(\boldsymbol\sigma, \boldsymbol\sigma'),
\]
 so the algebra does close, as it is to be expected from counting degrees of freedom.

As a consequence the elements $^\ast H_\tau[\xi]$, together with
$G_\tau [\bar\alpha]:= \int d\boldsymbol\sigma
\:\bar\alpha(X(\boldsymbol\sigma)) G(\boldsymbol\sigma)$ and
$C_\tau[\bar\beta]:=\int d\boldsymbol\sigma
\:\bar\beta(X(\boldsymbol\sigma)) C(\boldsymbol\sigma)$, form a
closed algebra under the Poisson brackets.

On the basis of the above discussion let us now derive explicit
expressions for the generators of the Lie algebra of space-time
dif\/feomorphisms associated with the anti-homomorphism
(\ref{antihomo}) and investigate whether these Lie algebra can be
extended with the smeared elements $G_\tau[\bar\alpha]$ and~$C_\tau[\bar\beta]$ and, if so what would be the interpretation of
such an extension. For this purpose let us f\/irst begin by deriving
the Poisson bracket of the projection $A_a$ of the 4-vector
potential f\/ield $A_\alpha$ on the hypersurface $\Sigma$ with $^\ast
H_\tau[\xi]$. Making use of (\ref{Cons1}) and (\ref{modsupham}) we
get
\begin{gather}
\{ A_a(\boldsymbol\sigma) , {^\ast H}_\tau[\xi]\}= \int d\boldsymbol\sigma' \: \{ A_a(\boldsymbol\sigma) ,
-\xi^\alpha(\boldsymbol\sigma') n_\alpha(\boldsymbol\sigma') {^\ast{{\mathcal H}}_{\perp}} +
\xi^\alpha {X_\alpha}^b (\boldsymbol\sigma')\:{^\ast{{\mathcal H}}_{b}} \}\nonumber\\
\phantom{\{ A_a(\boldsymbol\sigma) , {^\ast H}_\tau[\xi]\}}{} = -\xi^\alpha n_\alpha \gamma^{-\frac{1}{2}} \gamma_{ab}\pi^b + (\xi^\alpha A_\alpha)_{,a} + \xi^\alpha {X_\alpha}^b F_{ba}
= (\pounds_{\boldsymbol\xi} A_\beta) {X^\beta}_{a},\label{max5}
\end{gather}
after also making use of the expression
\begin{gather*}
\pi^a :=\frac{\delta{\mathcal L}}{\delta \dot A_a(\boldsymbol\sigma)} =-\gamma^{\frac{1}{2}}
\gamma^{ab}F_{\perp b},
\end{gather*}
for the momentum canonical conjugate to $A_a$ (c.f.\ equation~(3.10) in \cite{Kuchar2}).
Now, since the right side of (\ref{max5}) represents another gauge vector potential on $\Sigma$, it clearly follows
that
\begin{gather*}
\{\{ A_a(\boldsymbol\sigma) , {^\ast H}_\tau[\xi]\}, {^\ast H}_\tau[\eta]\}= (\pounds_{\boldsymbol\eta}\pounds_{\boldsymbol\xi} A_\beta) {X^\beta}_{a},
\end{gather*}
and interchanging the symbols $\boldsymbol\xi$, $\boldsymbol\eta$ on the left side above, substracting and using the Jacobi identity, yields
\begin{gather*}
\{ A_a(\boldsymbol\sigma) ,\{ {^\ast H}_\tau[\xi], {^\ast H}_\tau[\eta]\}\}= -(\pounds_{[\boldsymbol\xi,\boldsymbol\eta]} A_\beta) {X^\beta}_{a}.
\end{gather*}
We can therefore write the map
\begin{gather*}
\{ A_a(\boldsymbol\sigma) , {^\ast H}_\tau[\xi]\}\leadsto {^\ast \hat H}_\tau[\xi]\triangleright  A_a(\boldsymbol\sigma),
\end{gather*}
where
\begin{gather}\label{max9}
\delta_{\boldsymbol\xi}\equiv {^\ast \hat H}_\tau[\xi]:=({X^{\beta}}_a\circ \pounds_{\boldsymbol\xi})
\end{gather}
is a derivation operator which when acting on a 4-vector potential $A_\beta$ it projects its Lie derivative onto the hypersurface $\Sigma$.

Consider next the Poisson bracket of the scalar f\/ield $\psi$ with $^\ast H_\tau[\xi]$. Again, from~(\ref{Cons1}) and~(\ref{modsupham}) we get
\begin{gather}\label{max10}
\{ \psi(\boldsymbol\sigma), {^\ast H}_\tau[\xi]  \}=[\xi^\alpha (-n_\alpha \gamma^{-\frac{1}{2}} {\pi^a}_{,a}
+ {X_\alpha}^a \psi_{,a})](\boldsymbol\sigma).
\end{gather}
Similarly for the time evolution of $\psi$, derived from the total Hamiltonian, we obtain
\begin{gather}\label{max11}
\dot\psi= \{ \psi(\boldsymbol\sigma),\int d\boldsymbol\sigma' \:(N \:{^\ast{{\mathcal H}}_{\perp}}(\boldsymbol\sigma')
+ N^a \:{^\ast{{\mathcal H}}_{a}}(\boldsymbol\sigma')) \}=N \gamma^{-\frac{1}{2}} {\pi^a}_{,a} + N^a \psi_{,a}.
\end{gather}
Moreover, since
\begin{gather*}
\dot\psi :=\frac{\partial X^\alpha}{\partial\tau}\psi_{,\alpha}=N^\alpha \psi_{,\alpha} =N^\alpha (n_\alpha \psi_{,\perp} +{X_\alpha}^a \psi_{,a})
=-N\psi_{,\perp} + N^a \psi_{,a},
\end{gather*}
which when substituted into (\ref{max11}) implies that
$\psi_{,\perp}=-\gamma^{-\frac{1}{2}} {\pi^a}_{,a}$, and hence (from
(\ref{max10})) that \begin{gather*}
\{ \psi(\boldsymbol\sigma),
{^\ast H}_\tau[\xi]  \}=(\xi^\alpha
\psi_{,\alpha})(\boldsymbol\sigma)
=\pounds_{\boldsymbol\xi}\psi(\boldsymbol\sigma). \end{gather*} It clearly
follows from this that \begin{gather*}
\{ \psi(\boldsymbol\sigma),
\{{^\ast H}_\tau[\xi], {^\ast H}_\tau[\eta]
\}\}=-\pounds_{[\boldsymbol\xi,\boldsymbol\eta]}\psi(\boldsymbol\sigma)
\end{gather*} so for the action of ${^\ast H}_\tau[\xi]$ on scalar f\/ields we
can therefore also write the morphism (\ref{max9}), $ {^\ast
H}_\tau[\xi]\leadsto\delta_{\boldsymbol\xi}\equiv {^\ast \hat
H}_\tau[\xi]:=({X^{\beta}}_a\circ \pounds_{\boldsymbol\xi}),$
provided it is naturally understood that the surface projection
${X^{\beta}}_a$ acts as an identity on scalars. It should be clear
from the above analysis that these derivations
$\delta_{\boldsymbol\xi}$, as def\/ined in (\ref{max9}), are indeed
full space-time dif\/feomorphisms.

Let us now turn to the elements $G(\boldsymbol\sigma)$ and $C(\boldsymbol\sigma)$ of the algebra of constraints (\ref{diracb2}). The Poisson algebra
of the mapping
 $\bar\alpha\to G_{\tau[\bar\alpha]}= \int_{\Sigma} d{\boldsymbol\sigma'}\ \bar\alpha(X(\boldsymbol\sigma')) G(\boldsymbol\sigma')$,
with $A_a$ is
\begin{gather}\label{pal1}
\{A_a (\boldsymbol\sigma), G_\tau[\bar\alpha]\}=-\partial_a \bar\alpha,
\end{gather}
 and, making use of (\ref{max5}), we get
\begin{gather}\label{pal2}
\{\{A_a (\boldsymbol\sigma), G_\tau[\bar\alpha]\},{^\ast H}_\tau[\xi]\} =-(\pounds_{\boldsymbol\xi}\partial_\beta\bar\alpha){X^\beta}_a.
\end{gather}
 Inverting the ordering of the constraints in the above brackets we also have
\begin{gather}\label{pal3}
\{\{A_a (\boldsymbol\sigma), {^\ast H}_\tau[\xi]\},G_\tau[\bar\alpha]\}=-\{(\pounds_{\boldsymbol\xi}A_\beta){X^\beta}_a ,G_\tau[\bar\alpha]\}
=  - \partial_{a} (\xi^{c} \partial_{c} \bar\alpha).
\end{gather}
Subtracting now (\ref{pal2}) from (\ref{pal3}), and making use of the Jacobi identity on the left side of the equation, results in
\begin{gather}\label{pal4}
\{A_a (\boldsymbol\sigma),\{ {^\ast H}_\tau[\xi],G_\tau[\bar\alpha]\}\}= \partial_a (\xi^\perp \bar\alpha_{,\perp}).
\end{gather}
Note that we could equally well have gotten this result by identifying $G_\tau[\bar\alpha]$ with a derivation through the map
\begin{gather}\label{pal3b}
G_\tau[\bar\alpha]\leadsto \hat G_\tau[\bar\alpha]:= -\int_{\Sigma} d\boldsymbol\sigma' (\partial_b \bar\alpha)(\boldsymbol\sigma') \frac{\delta}{\delta A_b(\boldsymbol\sigma') },
\end{gather}
which could be seen as resulting from integrating the smeared constraint by parts and identifying the canonical momentum $\pi^b$
with the functional derivative: $\pi^b\leadsto\hat\pi^b:=\frac{\delta}{\delta A_b(\boldsymbol\sigma')}$. Indeed,
acting f\/irst on $A_a$ with the derivation operator (\ref{max9}) gives
\begin{gather*}
\delta_{\boldsymbol\xi}\triangleright A_a \equiv  {^\ast \hat H}_\tau[\xi]\triangleright A_a:=({X^{\beta}}_a\circ \pounds_{\boldsymbol\xi}) A_a\nonumber\\
\phantom{\delta_{\boldsymbol\xi}\triangleright A_a}{} = -\xi^\alpha n_\alpha \gamma^{-\frac{1}{2}} \gamma_{ab}\pi^b + (\xi^\alpha n_\alpha A_\perp)_{,a}
+ (\xi^c)_{,a} A_c + \xi^c A_{a,c} ,
\end{gather*}
which, when followed by the action of (\ref{pal3b}) results in
\begin{gather}
\hat G_\tau[\bar\alpha]\triangleright(\delta_{\boldsymbol\xi}\triangleright A_a)  =
-\int_{\Sigma} d\boldsymbol\sigma' (\partial_b \bar\alpha)(\boldsymbol\sigma') \frac{\delta}{\delta A_b(\boldsymbol\sigma') }
\left((\xi^\alpha n_\alpha A_\perp)_{,a}
+ (\xi^c)_{,a} A_c + \xi^c A_{a,c} \right)(\boldsymbol\sigma)\nonumber\\
\phantom{\hat G_\tau[\bar\alpha]\triangleright(\delta_{\boldsymbol\xi}\triangleright A_a)}{}  =  -\partial_a (\xi^c \partial_c \bar\alpha).\label{pal3d}
\end{gather}
Alternating the order of the above derivations, a similar calculation gives
\begin{gather*}
\delta_{\boldsymbol\xi}\triangleright (\hat G_\tau[\bar\alpha]\triangleright A_a) =-\delta_{\boldsymbol\xi}\triangleright \partial_a \bar\alpha
=\pounds_{\boldsymbol\xi} (\partial_\beta \:\bar\alpha) {X^\beta}_a= -\partial_a (\xi^\gamma \partial_\gamma \bar\alpha),
\end{gather*}
and subtracting from this (\ref{pal3d}) yields
\begin{gather*}
[\delta_{\boldsymbol\xi} , \hat G_\tau[\bar\alpha] ]\triangleright A_a = -\partial_a (\xi^\perp \partial_\perp \bar\alpha),
\end{gather*}
which could be thought to imply an algebra homomorphism when compared with (\ref{pal4}).
Observe, however, that if we evaluate  the Poisson bracket of $ {^\ast H}_\tau[\xi]$ and $ G_\tau[\bar\alpha]$ directly from (\ref{Cons1}) and (\ref{modsupham})
we get
\begin{gather}
\{ {^\ast H}_\tau[\xi],G_\tau[\bar\alpha]\} = \int_{\Sigma} d\boldsymbol\sigma \left(-\xi^\perp \bar\alpha_{,\perp}G +
\xi^\perp (\partial_a \bar\alpha) \gamma^{\frac{1}{2}}\gamma^{ab} C_{,b}  \right)(\boldsymbol\sigma)\nonumber\\
\phantom{\{ {^\ast H}_\tau[\xi],G_\tau[\bar\alpha]\}}{}  =  -G_{\tau}[\xi^\perp \bar\alpha_{,\perp}] - C_{\tau}[(\xi^\perp (\partial_a \bar\alpha) \gamma^{\frac{1}{2}}\gamma^{ab})_{,b}].\label{pal5}
\end{gather}
This result remains
compatible with (\ref{pal4}) because $C(\boldsymbol\sigma)$ acts as a projector when operating on the gauge vector f\/ield $A_a$.
But, because the right hand side of the equation contains a linear combination of the smeared constraints $G_\tau$ and $C_\tau$,
there is no way that we could implement the mapping (\ref{pal3b}) to get an homomorphism between the Poisson bracket (\ref{pal5})
and the Lie bracket $[\delta_{\boldsymbol\xi} , \hat G_\tau[\bar\alpha] ]$, as may be easily seen in fact when calculating the
later with (\ref{max9}) and (\ref{pal3b}).

Similarly, if we now consider the Poisson bracket of the map $\bar\beta{\to} C_{\tau}[\bar\beta]{=}
\int_{\Sigma} d{\boldsymbol\sigma'}\bar\beta(X(\boldsymbol\sigma')) C(\boldsymbol\sigma')$
with ${^\ast H}_\tau[\xi]$ we f\/ind (again making use of (\ref{Cons1}) and (\ref{modsupham})) that
\begin{gather}
\{C_\tau[\bar\beta], {^\ast H}_\tau[\xi]\} = \int_{\Sigma} d\boldsymbol\sigma [\xi^\alpha \bar\beta_{,\alpha}C + \xi^\perp \bar\beta\gamma^{-\frac{1}{2}}G
+\xi^a \bar\beta C_{,a}](\boldsymbol\sigma)\nonumber\\
\phantom{\{C_\tau[\bar\beta], {^\ast H}_\tau[\xi]\}}{}  =  C_\tau [\xi^\alpha \bar\beta_{,\alpha}-(\xi^a \bar\beta)_{,a}] + G_\tau [\xi^\perp \bar\beta\gamma^{-\frac{1}{2}}].\label{pal10}
\end{gather}
However, if we were to assume valid the derivation operator map $C_\tau[\bar\beta]\leadsto \hat C_\tau[\bar\beta]
= \int_{\Sigma} d\boldsymbol\sigma \bar\beta\frac{\delta}{\delta\pi(\boldsymbol\sigma )}$, it would then clearly follow
that
\begin{gather*}
[\delta_{\boldsymbol\xi} , \hat C_\tau[\bar\beta] ]\triangleright A_a=0.
\end{gather*}
This result immediately enters into conf\/lict with (\ref{pal10}), where such a morphism of algebras, involving $\hat C_\tau[\bar\beta]$ together with (\ref{pal3b}), would yield
\begin{gather*}
\{A_a, \{C_\tau[\bar\beta], {^\ast H}_\tau[\xi]\}\}\leadsto [\delta_{\boldsymbol\xi} , \hat C_\tau[\bar\beta] ]\triangleright A_a
=-\partial_a(\xi^\perp \bar\beta\gamma^{-\frac{1}{2}}).
\end{gather*}

Consequently, the largest Lie algebra that we can associate with the
Poisson algebra (\ref{diracb2}) is the one of space-time
dif\/feomorphisms, given by the homomorphism implied by
(\ref{antihomo}) and originating from the sub-algebra of the
super-Hamiltonian and super-momenta described by the f\/irst 3
equations in (\ref{diracb2}). We shall return to this observation
later on, as it is essential for our conclusions. First we need
however to relate our results derived so far with some basic aspects
of gauge theory as formulated from the point of view of principal
f\/iber bundles.

\section{Gauge transformations}\label{sec4}

Recall (c.f.\ e.g.~\cite{bleecker}) that a gauge transformation of
a principal f\/iber bundle (PFB) $\pi:P\to {\mathcal M}$, with
structure Lie group $\mathcal G$, is an automorphism $f:P \to P$
such that $f(pg)= f(p)g$ and the induced dif\/feomorphism $\bar f :
{\mathcal M}\to {\mathcal M}$, def\/ined by $\bar
f(\pi(p))=\pi(f(p))$, is the identity map ${\bar f}=1_{\mathcal M}$
(i.e.\ $\pi(p)= \pi(f(p)))$. Moreover, if we def\/ine $f:P\to P$
by $f(p)= p\zeta(p)$, where $\zeta$~is an element of the space
$C(P,\mathcal G)$ of all maps such that
$\zeta(pg)=g^{-1}\cdot\zeta(p)={\rm Ad}_{g^{-1}} \zeta(p)$ (so $\mathcal G$
acts on itself by an adjoint action), then $C(P,\mathcal G)$ is
naturally anti-isomorphic to the group of gauge transformations
$GA(P)$. That is,
for $f, f^{\prime} \in GA(P)$ and $\zeta, \zeta'\in C(P,\mathcal G) $ we have that $(f\circ f^\prime)(p)=p(\zeta'(p)\zeta(p))$.

From the above, it can be readily shown that \begin{gather*}
f_{\ast}(\sigma_{u\ast}{\bf
X})=\frac{d}{dt}\left(R_{\zeta(p)^{-1}\circ \zeta(\sigma_u
(\gamma(t)))}f(p)\right)|_{t=0} + R_{\zeta(p)\ast}(\sigma_{u\ast}{\bf
X}), \end{gather*} where ${\bf X}\in T{\mathcal M}$ or, writing
$\zeta(p)^{-1}\circ \zeta(\sigma_{u} (\gamma(t))):= e^{t\frak b}$ as
an element of a one-parameter subgroup of $\frak G$,
\begin{gather*}
f_{\ast}(\sigma_{u\ast}{\bf X})= {\frak b}^{\ast}_{f(p)} + R_{\zeta(p)\ast}(\sigma_{u\ast}{\bf X}),
\end{gather*}
where ${\frak b}^{\ast}_{f(p)}$ is the fundamental vector f\/ield on $f(p)$ corresponding
to
\begin{gather}\label{gauge3.4}
{\frak b}=L^{-1}_{\zeta(p)\ast}\zeta_{\ast}(\sigma_{u}\ast {\bf X} ).
\end{gather}
Consequently,
\begin{gather}\label{gauge3}
(\sigma_u^\ast f^\ast \omega)({\bf
X}) = {\frak b}
+ Ad_{(\sigma_u^\ast\zeta)(X)^{-1}} (\sigma_u^\ast \omega)({\bf X}).
\end{gather}
In the above expressions, $\omega_{f(p)}$ is a connection 1-form at $f(p)\in P$, $(f^\ast \omega)_p$
is its pull-back to~$p$ with the gauge map $f$ and $(\sigma_u^\ast f^\ast \omega)_{\pi(p)}$
is in turn its pull-back with the local section $\sigma_u$ to \mbox{a~1-form} on ${\mathcal U}\subset {\mathcal M}$, the map
$\gamma:{\Bbb R} \to {\mathcal U}$ is a curve in the base manifold with $\frac{d}{dt}\gamma(t)|_{t=0}= \bf X$, and
$(\sigma_u^\ast\zeta)(X^\mu)$ is a space-time-valued element of $\mathcal G$.

Write now $\zeta$ as an element of a one-parameter subgroup of
$C(P,\mathcal G)$ by means of the exponential map \begin{gather}\label{expmap}
\zeta = \exp(-t\alpha^{B} T_B), \end{gather} where $\alpha^{B} T_B
:=\boldsymbol\alpha$ is an element of the gauge algebra space
$C(P,\frak g)$, and the $T_B$ denote the basis matrices of the Lie
algebra $\frak g$ associated with $\mathcal G $. Replacing
(\ref{expmap}) into (\ref{gauge3.4}) and (\ref{gauge3}) we get{\samepage
\begin{gather}
(\sigma_u^\ast ({ R}_{\exp(-t\alpha^{B}(p) T_B)})^\ast \omega)({\bf X}) =
\frac{d}{ds}[\exp(t\bar\alpha^{B}(X) T_B) \exp(-s\bar\alpha^{B}(\gamma(s)) T_B)]|_{s=0}\nonumber\\
\phantom{(\sigma_u^\ast ({ R}_{\exp(-t\alpha^{B}(p) T_B)})^\ast \omega)({\bf X}) =}{}+ {\rm Ad}_{\exp(t\bar\alpha^{B}(X) T_B)} (\sigma_u^\ast \omega)({\bf X}),\label{gauge4}
\end{gather}
where} $\bar\alpha^B:= (\sigma_u^\ast \alpha^B)$.
The inf\/initesimal version of (\ref{gauge4}) follows directly by dif\/ferentiating both sides
of the above equation with respect
to the parameter $t$ and evaluating at zero. We therefore arrive at
\begin{gather}\label{gauge5}
\delta_{\bar\alpha} A :=\frac{d}{dt} (\sigma_u^\ast ({ R}_{\exp(-t\alpha^{B} T_B)})^\ast \omega)|_{t=0}
=-d\bar{\boldsymbol\alpha} -[A, \bar{\boldsymbol\alpha}] = -D\bar{\boldsymbol\alpha} \in \bar\Lambda^1({\mathcal M}, {\frak g}) ,
\end{gather}
where $\Lambda^1({\mathcal M}, {\frak g})$ denotes the space of 1-forms on ${\mathcal M}$
valued in the Lie algebra ${\frak g}$.

Making use of (\ref{gauge5}) in the expression for the Yang--Mills curvature:
\begin{gather*}
F:=DA= dA + \frac{1}{2} [A, A],
\end{gather*}
we obtain that
\begin{gather}\label{vymf}
\delta_{\bar\alpha} F = [\bar{\boldsymbol\alpha}, F].
\end{gather}
In the particular case
where the one-parameter group is Abelian, it immediately follows that~(\ref{gauge5}) and~(\ref{vymf}) simplify to
\begin{gather}\label{gauge6}
\delta_{\bar\alpha} A = -i d\bar\alpha,
\end{gather}
and
\begin{gather*}
\delta_{\bar\alpha} F = 0.
\end{gather*}
This last result merely states the well know fact that the electromagnetic f\/ield strength is gauge independent
(i.e.\ it is independent of the choice of local trivialization).

Moreover, since (\ref{gauge6}) implies that $\delta_{\bar\alpha}
A_\mu = -i \partial_\mu\bar\alpha$, we obtain, by projecting on
the sheet  $\Sigma$  with ${X^\mu}_a$, \begin{gather}\label{gauge7}
\delta_{\bar\alpha} A_a =-i\ \partial_a
\bar\alpha(X(\boldsymbol\sigma)). \end{gather} Let us now turn to the Gauss
constraint $G(\boldsymbol\sigma)$, introduced in (\ref{Cons1}), and
to the smearing map \begin{gather}\label{mapgauss} \bar\alpha\to
G_{\tau[\bar\alpha]}= \int_{\Sigma} d{\boldsymbol\sigma'}
\bar\alpha(X(\boldsymbol\sigma'))  G(\boldsymbol\sigma'). \end{gather}
Comparing (\ref{pal1}) with (\ref{gauge7}) we see that
\begin{gather}\label{gauss4} i\{A_a ,
G_\tau[\bar\alpha]\}\cong\delta_{\bar\alpha}A_a, \end{gather} so the Poisson
bracket of the projection  $A_a$  of the gauge 4-vector on the
space-like hypersurface~$\Sigma$  with
 the Gauss constraint smeared with the scalar function $\bar\alpha(X(\boldsymbol\sigma'))$ is the same as
the pullback to $\mathcal M$ of the inf\/initesimal action of the gauge algebra of the PFB with group $U(1)$ on
the connection one-form $\omega$ (c.f.\ equation~(\ref{gauge5})) evaluated on a tangent vector to $\Sigma$.

In addition, for $f\in GA(P)$, it is a simple matter to show that if $\omega$ is a connection 1-form
then the pullback $f^\ast \omega$ is also a connection 1-form. This theorem follows immediately
by noting f\/irst that the action of $f^\ast \omega$ on a fundamental vector yields its corresponding
Lie algebra generator, and second that the requirement $\omega_{pg}(R_{g\ast} X)= {\rm Ad}_{g^{-1}}\omega_p (X)$
in the def\/inition of a connection 1-form is directly satisf\/ied when acting on $\omega$ with the pullback of $f\circ R_g = R_g \circ f$, which in turn
is equivalent the automorphism condition $f(pg)= f(p)g$.

Let now $V$ be a vector space on which $\mathcal G$ acts from the left. If $L_g :V\to V$ is linear, then
the homomorphism ${\mathcal G}\to {\mathcal GL}(V)$ by $g \mapsto L_g$ is a representation of $\mathcal G$.
In this case $C(P,V)$ will denote the space of all maps $\zeta: P\to V$ such that $\zeta(pg)=g^{-1}\cdot \tau(p)$
and the elements of $C(P,V)$ correspond to particle f\/ields.

In particular,  $C(P,V)=\bar\Lambda^0(P,V)$, where, in general, $\bar\Lambda^k(P,V)$ is the space of $V$-valued
dif\/fe\-ren\-tial $k$-forms $\varphi$ on $P$ such that
\begin{gather*}
 R^{\ast}_g \varphi = g^{-1}\cdot \varphi,\nonumber\\
 \varphi({\bf Y}_{1},\dots, {\bf Y}_{k})=0, \quad \text {if any one of the} \  {\bf Y}_1,\dots {\bf Y}_k \in T_p P  \ \text {is vertical}.
\end{gather*}
Making now use of the exponential map (\ref{expmap}) it readily follows that
\begin{gather}\label{gauge9}
f^{\ast} \varphi= \zeta^{-1}\cdot\varphi.
\end{gather}

Or, dif\/ferentiating  with respect to $t$ and evaluating at $t=0$, we arrive at the following
inf\/initesimal version of (\ref{gauge9}):
\begin{gather}\label{gaugefield}
\delta_{\bar\alpha}\bar\varphi= \bar\alpha^{B} T_B \cdot\bar\varphi.
\end{gather}
Furthermore, related to our discussion in the following sections, note that from the def\/inition of dif\/feomorphisms
we have that $R_{g} \circ f = f\circ R_{g}$, thus acting with the pull-back of this equality on any element $\kappa \in \bar\Lambda^{k}(P,V)$,
 and recalling that the action of the dif\/ferential $f\ast$ on a fundamental f\/ield
$B^{\ast}$ is a fundamental f\/ield, it then immediately follows that
$(f^\ast \kappa)(B^\ast)= \kappa(B^\ast)=0$. Hence $f^{\ast}
\kappa\in \Lambda^{k}(P,V)$, $k=0,1,2\dots$, and since
$C(P,V)=\bar\Lambda^{0} (P,V)$ it also follows that the gauge group
$GA(P)$ acts on particle f\/ields via pull-back, so that
\begin{gather}\label{partf2} f^{\ast} \varphi(p)= \varphi(f(p)), \end{gather} i.e.\
if $\varphi$ is a particle f\/ield, so is also $f^{\ast} \varphi$.

Using the above results we can now formulate the multiplication rules for gauge and particle f\/ields
under gauge transformations, when pulled-back to the base space $\mathcal M$. Thus, given  two
$\frak g$-valued potential 1-forms $A, A'\in \Lambda^1({\mathcal M}, \frak g)$, their product is def\/ined by
\begin{gather*}
[A, A']:= \big(A^a\wedge A'^b \big)\otimes [T_a, T_b],
\end{gather*}
while the product of two particle f\/ields $\varphi_1, \varphi_2 \in C(P,V)$ is by simple point multiplication.
Now, as shown previously, the action of an element $f\in GA(P)$ on a connection 1-form and on a particle f\/ield is
via pull-back (c.f.\ equations~(\ref{gauge3}) and (\ref{partf2})) and since the pull-back of a~connection is a connection
and the pull-back of a particle f\/ield is a particle f\/ield, it therefore follows that
\begin{gather*}
f:[A, A'] \leadsto [(\sigma_u^\ast f^\ast \omega_1), (\sigma_u^\ast f^\ast \omega_2)],
\\
f: (\sigma^\ast_u\varphi_1)(\pi(p))\cdot (\sigma^\ast_u\varphi_2)(\pi(p))\leadsto (\sigma_u^\ast f^\ast\varphi_1)(\pi(p))\cdot
(\sigma_u^\ast f^\ast\varphi_2)(\pi(p)).
\end{gather*}
By (\ref{gauge5}) and (\ref{gaugefield}), the inf\/initesimal expression for the above is:
\begin{gather}
\delta_{\bar\alpha}\left([A, A']({\bf X}_1 ,{\bf X}_2 )\right)\nonumber\\
\qquad{}:=\mu \big[(\delta_{\bar\alpha}\otimes 1 +1\otimes \delta_{\bar\alpha})
\big(A^a({\bf X}_1)\otimes A'^b({\bf X}_2)-
A^a({\bf X}_2)\otimes A'^b({\bf X}_1)\big)\big]\otimes [T_a, T_b]\nonumber\\
\qquad{}=(\delta_{\bar\alpha}A^a \wedge A'^b -A^a \wedge \delta_{\bar\alpha}A'^b)({\bf X}_1 ,{\bf X}_2 )\otimes [T_a, T_b],\label{defvwmult}
\end{gather}
and
\begin{gather}\label{multpf}
\delta_{\bar\alpha}\left(\bar\varphi_1 (\pi(p))\cdot \bar\varphi_2 (\pi(p)) \right) = \delta_{\bar\alpha}(\bar\varphi_1(\pi(p)))\cdot
\bar\varphi_2(\pi(p)) + \bar\varphi_1(\pi(p))\cdot \delta_{\bar\alpha}(\bar\varphi_2(\pi(p))),
\end{gather}
respectively. This last result implies that under an inf\/initesimal gauge transformation the pro\-duct of two particle f\/ields
transforms according to the Leibniz rule. We can therefore give this inf\/initesimal transformations the structure of a Hopf
algebra with coproduct $\Delta \delta_{\bar\alpha} =\delta_{\bar\alpha} \otimes 1 +1\otimes \delta_{\bar\alpha}$, so that
\begin{gather*}
\delta_{\bar\alpha}\left(\bar\varphi_1 (\pi(p))\cdot \bar\varphi_2 (\pi(p)) \right) =\mu[\Delta \delta_{\bar\alpha}
\left(\bar\varphi_1 (\pi(p))\cdot \bar\varphi_2 (\pi(p)) \right) ].
\end{gather*}

From the above discussion we can derive some additional insight into
the implications of the PFB  point  of view of gauge transformations
on our previous results. We thus see that since gauge
transformations are automorphisms on the f\/ibers that project to the
identity on the base space, the Gauss constrain -- which we have seen
here to be related to the pull-back of the inf\/initesimal gauge
transformations, and which was shown in Section~\ref{sec3} to be needed in order
to close the algebra in (\ref{algebra1.1}) -- occurs in the extended
algebra (\ref{diracb2}) primarily as part of the super-Hamiltonian
and super-momenta associated with the Lie algebra of space-time
dif\/feomorphisms. Its independent appearance is then only as a
constraint which, together with $C(\boldsymbol\sigma)\simeq 0$, have
to be implemented at the end as strong conditions in order to
recover the Maxwell theory. This provides an additional natural
explanation for why these two constraints can not be mapped into
derivations that could lead to an enlarged Lie algebra beyond the
one of the space-time dif\/feomorphisms.

\section{Noncommutative gauge theories}\label{sec5}

With these results in hand, let us now consider an approach for
extending the theory of~gauge f\/ields to the noncommutative space-time
case, by specif\/ically concentrating on the vacuum Maxwell f\/ield
discussed in the last two sections, and by following the procedure
introduced in~\cite{nos1}. Recall, in particular, that~-- because of
the anti-homomorphism that can be established between the Poisson
sub-algebra of the constraints occurring in the f\/irst 3 lines of
(\ref{diracb2}), for the modif\/ied theory in extended phase space,
and the Lie algebra $\pounds\,\text{dif\/f} \, \mathcal M$~-- we can use
the latter to investigate the deformed space-time isometries of the
system by requiring that this sub-algebra of constraints, modif\/ied
by the noncommutativity of space-time, should continue obeying the
Dirac relations, relative to the Dirac brackets resulting from
admitting an arbitrary symplectic structure in the action
(\ref{Nocom1}). This, as shown in \cite{nos1}, was needed in turn in
order to incorporate into the parametrized canonical formalism the
dynamical origin of star-noncommutativity from quantum mechanics \cite{ros1}.
Moreover, since the constraints depend on the metric of the
embedding space-time, this last step would require in general a well
developed theory of quantum mechanics in curved spaces and knowledge
of the commutators of the operators representing the phase space
coordinates. We shall defer such more general considerations for
some future presentation, and concentrate here only on the case of
f\/ields on f\/lat Minkowski space-time and the corresponding quantum
mechanics for the extended Weyl--Heisenberg group.

Consequently, admitting  a symplectic structure in the action
(\ref{modelect}) we have
\begin{gather*}
    S[z]=\int d^4\sigma  \big(\mathcal {B}(z)_{A} {\dot z}^A   -N^{\alpha}
(^\ast\tilde{\mathcal H}_{\alpha})  -
M {G}(\boldsymbol\sigma) -T C(\boldsymbol\sigma) \big),
\end{gather*}
with the symplectic variables $z^A=(X^\alpha, A_a, \psi; P_\alpha,
\pi^a, \pi)$ and symplectic potentials $\mathcal{B}(z)_A$ to be
determined by a prescribed symplectic structure. Here $M$, $T$ are the
additional Lagrange multipliers needed to recover Maxwell's
electrodynamics and the tildes on the constraints needed of the
formerly introduced quantities, in order that their Dirac-bracket
algebra originated by the new symplectic structure is identical to
their sub-algebra in (\ref{diracb2}). That is, we want to maintain
the algebra of these constraints invariant by utilizing new twisted
generators. (Observe however, that since the
${G}(\boldsymbol\sigma)$ and $ C(\boldsymbol\sigma)$ can not form
part of our Lie algebra of space-time isometries, but are strictly
constraints to be implemented in order to retrieve
Maxwell's electromagnetism, their action on gauge and particle f\/ields will be determined by the arguments given at the end of this section.)

As noted in \cite{nos1}, the symplectic structure is def\/ined by,
\begin{gather}\label{seccom}
\omega_{AB}:=\frac{\partial \mathcal{B}_B}{\partial z^A}
-\frac{\partial \mathcal{B}_A}{\partial z^B},
\end{gather}
from where we can readily solve for the symplectic potentials, which
are def\/ined up to a canonical transformation. The resulting
second-class constraints can then be eliminated by introducing Dirac
brackets, according to a scheme analogous to the one described in
the above cited paper, from where the inverse of the symplectic
structure is additionally def\/ined through the Dirac-brackets for the
symplectic variables $z^A$. Hence the Dirac brackets for the
symplectic variables are given by \begin{gather}\label{dirbrack} \{ z^A, z^B
\}^\ast := \{ z^A, z^B \} -\{ z^A, \chi_C \}\:\omega^{CD} \{ \chi_D,
z^B \}=\omega^{AB}, \end{gather} where $\chi_A =\pi_{z^A}
-\mathcal{B}(z)_A\simeq 0 $ are the second-class constraints. More
specif\/ically, based on the premise that quantum mechanics is a
minisuperspace of f\/ield theory and for a quantum mechanics on f\/lat
Minkowski space-time based on the extended Weyl--Heisenberg group, we
have shown in \cite{ros1} that the WWGM formalism implies that, for
the phase space variables to have a~dynamical character, we need to
modify their algebra by twisting their product according to \begin{gather} \mu
(X^\alpha \otimes X^\beta)\leadsto\mu_\theta (X^\alpha \otimes
X^\beta):= X^\alpha (\tau,\boldsymbol\sigma) \star_{\theta} X^\beta
(\tau,\boldsymbol\sigma'),\label{4.4} \end{gather} where \begin{gather}\label{starprod}
\star_{\theta}:= \exp\left[\frac{i}{2}\theta^{\mu\nu}\int
d\sigma''\frac{\overleftarrow\delta} {\delta
X^{\mu}(\tau,\boldsymbol\sigma'')}\frac{\overrightarrow\delta}{\delta
X^{\nu}(\tau,\boldsymbol\sigma'')}\right] , \end{gather} and where, since the
embedding space-time variables are functionals of the foliation, we
use functional derivatives. Also, since f\/ields are in turn functions
of the embedding space-time variables their multiplication in the
noncommutative case is inherited from (\ref{4.4}). Moreover, using
this $\star$-product we can now def\/ine the commutator
\begin{gather}
[X^\alpha (\tau,\boldsymbol\sigma), X^\beta (\tau,\boldsymbol\sigma')]_\theta:= X^\alpha (\tau,\boldsymbol\sigma)\star_\theta X^\beta (\tau,\boldsymbol\sigma')
-X^\beta (\tau,\boldsymbol\sigma')\star_\theta X^\alpha (\tau,\boldsymbol\sigma)\nonumber\\
\phantom{[X^\alpha (\tau,\boldsymbol\sigma), X^\beta (\tau,\boldsymbol\sigma')]_\theta}{} =i\theta^{\alpha\beta}\delta(\boldsymbol\sigma,\boldsymbol\sigma'),\label{starcomm}
\end{gather}
and let
\begin{gather*}
\{X^\alpha, X^\beta \}^\ast =[X^\alpha (\tau,\boldsymbol\sigma), X^\beta (\tau,\boldsymbol\sigma')]_{\star\theta}=
i\theta^{\alpha\beta}\delta(\boldsymbol\sigma,\boldsymbol\sigma').
\end{gather*}

On the other hand, def\/ining the map
\begin{gather}\label{dar1}
\tilde X^\alpha = X^\alpha + \frac{\theta^{\alpha\beta}}{2} P_\beta,
\end{gather}
it follows from (\ref{dirbrack}) that \begin{gather}\label{diracb3} \{\tilde
X^\alpha, \tilde X^\beta \}^\ast =0, \end{gather} and \begin{gather*}
\{ ^\ast{\tilde {\mathcal H}}_{\alpha}(\vec{\sigma}) , ^\ast{\tilde
{\mathcal H}}_{\beta}(\vec{\sigma'})\}^\ast=0. \end{gather*} Thus, in parallel
to (\ref{antihomo}), we have \begin{gather*}
\{ {^\ast}{\tilde
H}_\tau[\xi], {^\ast}{\tilde H}_\tau[\rho] \}^\ast=-{^\ast}{\tilde
H}_\tau[\pounds_{\xi}\rho]. \end{gather*} Furthermore, making the
identif\/ication $P_\beta = -i\frac{\delta}{\delta X^{\beta}}$ in the
Darboux map (\ref{dar1}) we can write \begin{gather}\label{starcomm3} \tilde
X^{\alpha} \leadsto \hat{\tilde X}^\alpha
=(X^\alpha)\:\star^{-1}_\theta := (X^\alpha)
\exp\left[-\frac{i}{2}\theta^{\mu\nu}\int
d\sigma''\frac{\overleftarrow\delta} {\delta
X^{\mu}(\tau,\boldsymbol\sigma'')}\frac{\overrightarrow\delta}{\delta
X^{\nu}(\tau,\boldsymbol\sigma'')}\right] , \end{gather} where the
bi-dif\/ferential acting from the right on the embedding coordinates
$X^\alpha$ is the inverse of (\ref{starprod}). Hence
\begin{gather*}
\{\tilde X^\alpha, \tilde X^\beta \}^\ast\cong
[\hat{\tilde X}^\alpha , \hat{\tilde X}^\beta ]_{\star_\theta} =
[X^\alpha , X^\beta]\:\star^{-1}_\theta =0, \end{gather*}
since under point multiplication the embedding coordinates commute. So the map (\ref{starcomm3}) retrie\-ves~(\ref{diracb3}).

In addition, since multiplication in the algebra of the operators $\hat{\tilde X}^\alpha$ is by the $\star_{\theta}$-product
we can generalize the last result to
\begin{gather*}
\{(\tilde X^\alpha)^m, (\tilde X^\beta)^n \}^\ast\cong [(\hat{\tilde X}^\alpha)^m _\star, (\hat{\tilde X}^\beta)^n_\star ]_{\star_\theta}
= [(X^\alpha)^m, (X^\beta)^n]\:\star^{-1}_\theta =0.
\end{gather*}

We can therefore conclude from the above that, when replacing the
functional dependence on the embedding variables in the constraints
in (\ref{diracb2}) by the ``tilde'' variables (\ref{dar1}) and the
point multiplication of f\/ields by their $\star$-product, the
functional form of their algebra is evidently preserved for the
noncommutative case. That is, \begin{gather}\label{diracb6} \{ {^\ast}{\tilde
H}_\tau[\xi] , {^\ast}{\tilde H}_\tau[\eta]\}^\ast\cong [^\ast{\hat
H}_{\tau}[\xi],\ ^\ast{\hat
H}_{\tau}[\eta]]_{\star}\:\star^{-1}_\theta, \end{gather} and
\begin{gather}\label{moddif} ^\ast{\hat H}_\tau[\xi]=\delta_\xi\leadsto
 {^\ast{\hat H}}_\tau[\xi]\:\star^{-1}_\theta =\delta^\star_\xi, \end{gather}
where the multiplication $\mu_\theta$ of the algebra of generators
of dif\/feomorphisms $\delta^\star_\xi\in \pounds\,\text {dif\/f} \,\mathcal M$
is via the $\star_\theta$-product.

Consequently, by using the example of a modif\/ied electromagnetism
within the context of canonical parametrized f\/ield theory,
 it was shown that, by including additional constraints, Maxwell's equations could be recovered as well
 as the possibility
of also establishing for gauge f\/ield theories the anti-homomorphism
between Dirac-brackets of the modif\/ied constraints and  space-time
dif\/feomorphisms. Furthermore using our previous results in
\cite{nos1} where it was shown that noncommutativity in f\/ield theory~-- manifested as the twisting of the algebra of f\/ields~-- has a
dynamical origin in the quantum mechanical mini-superspace which,
for f\/lat Minkowski space-time, is related to an extended
Weyl--Heisenberg group, and including these results into the
symplectic structure of the parametrized f\/ield theory then allowed
us to derive the deformed Lie algebra
of the noncommutative space-time dif\/feomorphisms, as shown by (\ref{diracb6}) and (\ref{moddif}) above.

Moreover, making use of (\ref{moddif}) we can summarize the action
of space-time dif\/feomorphisms on particle f\/ields associated with
gauge theories, and the transition of the theory to the
noncommutative space-time case by means of the following functorial
diagrams:
\newcommand{\End}{\operatorname{End}}
\begin{gather}\label{functor1}
\begin{CD}
{^\ast} H_\tau[\xi]\;&\in&\;{\mathcal V} @>\theta>>{\mathcal V}^\star&&
\hskip-6.5pc\ni {^\ast}{\tilde  H}_\tau[\xi]= \int
d\vec\sigma(\tilde\xi^\perp \ {^\ast\tilde{\mathcal H}}_\perp +
\tilde\xi^a \ ^\ast\tilde{\mathcal H}^a)\\
\mathcal{C}@.   @VVV        {\mathcal{C}}@VVV \\
{^\ast}\hat H_{\tau}[\xi]\;&\in&\;\hat{\mathcal V} @>{{\mathcal
C}(\theta)}>>\hat{\mathcal V}^\star\; &\ni {^\ast}{\hat H}_\tau[\xi]\:\star^{-1}_\theta
\equiv \delta^\star_\xi
\end{CD}
\end{gather}
(where $\mathcal V$ denotes the space of constraints satisfying the algebra (\ref{diracb2}), ${\mathcal V}^\star$
is the corresponding space of constraints for the space-time noncommutative case with the embedding coordinates
mapped according to (\ref{dar1}) and $\hat{\mathcal V}$ , $\hat{\mathcal V}^\star$ denote the spaces of the Lie algebra of
dif\/feomorphisms and their corresponding twisted form, respectively);
\begin{gather}\label{functor2}
\begin{CD}
\bar\varphi\;&\in&\;{\mathcal A}  @>{\delta_\xi}>> {\mathcal A}&\ni& \hskip-5pc \delta_\xi\triangleright\bar\varphi\\
\mathcal{D}@.  @VVV       {\mathcal{D}}@VVV \\
 \bar\varphi\;&\in&\;{\mathcal A}_\theta @>{{\mathcal
D}(\delta^\star_{\xi})}>> {\mathcal A}_\theta \;&\ni&\; \delta^\star_\xi
\triangleright\bar\varphi =\delta^{\star}_{\xi} \star_\theta\bar\varphi(X(\tau,\boldsymbol\sigma))
\end{CD}
\end{gather}
(here ${\mathcal A}\; \text {denotes the module algebra of particle f\/ields}\; \bar\varphi\in C({\mathcal M},V)\; \text {with point multiplication $\mu$} $
and ${\mathcal A}_\theta$ is its noncommutative twisting with $\star$-multiplication $\mu_\theta:=\mu\circ
e^{\frac{i}{2}\theta^{\mu\nu} \partial_{\mu}\otimes\partial_{\nu}}$).

It then follows from these two diagrams that
 \begin{gather}\label{functor3} \{\bar\varphi,\; {^\ast}\hat H_{\tau}[\xi]\}\cong
\delta_\xi\triangleright\bar\varphi\mapsto \delta^{\star}_{\xi}
\star_\theta\bar\varphi(X(\tau,\boldsymbol\sigma))={^\ast}\hat H_\tau[\xi]\triangleright\bar\varphi. \end{gather}

 Note that the diagrams (\ref{functor1}), (\ref{functor2}) and equation~(\ref{functor3}) provide an explicit expression
for the mappings $\delta_\rho \mapsto \delta^{\star}_{\rho} $, which
in turn imply \begin{gather*}
\left[\delta^{\star}_{\rho},
\delta^{\star}_{\eta}\right]_{\star\theta}=\delta^\star_{\pounds_\rho
\eta}, \end{gather*} and \begin{gather}\label{virared3} \delta^{\star}_{\rho}
\star_\theta (\bar\varphi_1\star_\theta \bar\varphi_2) =\delta_\rho
(\bar\varphi_1\star_\theta \bar\varphi_2), \end{gather} where $\bar\varphi_1,
\bar\varphi_2 \in {\mathcal A}_\theta$.

 Note also that the universal
envelopes $U(\hat{\mathcal V})$ and $U(\hat{\mathcal V}^\star)$ of
the derivations $\delta_\xi$ and twisted derivations
$\delta^{\star}_{\xi}$ can be given the structure of  Hopf algebras.
Thus, in particular, we can obtain an explicit expression for the
coproduct in $U(\hat{\mathcal V}^\star)$ by making use of the
duality between product and coproduct, followed by the application
of equation (\ref{virared3}). We get \begin{gather*}
\mu_{\theta}\circ\Delta (\delta^{\star}_{\rho})(\bar\varphi_1\otimes\bar \varphi_2)=
\delta^{\star}_{\rho} \star_\theta(\bar\varphi_1\star_\theta \bar\varphi_2)=\delta_{\rho} (\bar\varphi_1\star_\theta \bar\varphi_2)\nonumber\\
\qquad{} =\mu(\delta_{\rho} \otimes 1+1\otimes \delta_{\rho})
(e^{\frac{i}{2}\theta^{\mu\nu} \partial_{\mu}\otimes \partial_{\nu}} \bar\varphi_1\otimes \bar\varphi_2)\nonumber\\
\qquad{}=\sum_n \frac{1}{n!}\left(\frac{i}{2}\right)^n \theta^{\mu_1 \nu_1} \cdots\theta^{\mu_n \nu_n}
\left[(\delta^\star_\rho \star_\theta\partial_{\mu_1 \dots\mu_n} \bar\varphi_1) e^{-\frac{i}{2}\theta^{\mu\nu}
{\overleftarrow \partial}_{\mu} {\overrightarrow\partial}_{\nu}}
\star_\theta\partial_{\nu_1 \dots \nu_n} \bar\varphi_2\
\right.\nonumber\\
\left.\qquad\quad {}+(\partial_{\mu_1 \dots\mu_n} \bar\varphi_1) e^{-\frac{i}{2}\theta^{\mu\nu}{\overleftarrow \partial}_{\mu}{\overrightarrow \partial}_{\nu}}\star_\theta
(\delta^{\star}_\rho \star_\theta\partial_{\nu_1\dots \nu_n} \bar\varphi_2)\right] \nonumber\\
\qquad{}=\mu_{\theta}\circ \left [e^{-\frac{i}{2}\theta^{\mu\nu}
\partial_{\mu}\otimes \partial_{\nu}}(\delta^\star_{\rho}\otimes 1 + 1\otimes\delta^\star_{\rho})
e^{\frac{i}{2}\theta^{\mu\nu} \partial_{\mu}\otimes \partial_{\nu}}\right](\bar\varphi_1\otimes \bar\varphi_2).
\end{gather*}
This result compares with the Leibniz rule given in \cite{Wess1}.
Furthermore, if we let
${\mathcal F}=e^{-\frac{i}{2}\theta^{\mu\nu}\partial_{\mu}\otimes \partial_{\nu}}\in
U(\hat{\mathcal V})\otimes U(\hat{\mathcal V})$, and def\/ine
$\bar\varphi_1\star_\theta \bar\varphi_2=\mu_\theta (\bar\varphi_1\otimes \bar\varphi_2):=
\mu({\mathcal F}^{-1}\triangleright (\bar\varphi_1\otimes \bar\varphi_2))$,
we then have \cite{bloh,bloh+}:
\begin{gather}
\delta_{\rho}(\bar\varphi_1\star_\theta \bar\varphi_2) = \delta_{\rho}\triangleright
\mu({\mathcal F}^{-1}\triangleright (\bar\varphi_1\otimes \bar\varphi_2))=
\mu[(\Delta \delta_{\rho}){\mathcal F}^{-1}\triangleright (\bar\varphi_1\otimes \bar\varphi_2))]\nonumber\\
\phantom{\delta_{\rho}(\bar\varphi_1\star_\theta \bar\varphi_2)}{} = \mu {\mathcal F}^{-1}[({\mathcal F}(\Delta \delta_{\rho}) {\mathcal F}^{-1})((\bar\varphi_1\otimes \bar\varphi_2)))]\nonumber\\
\phantom{\delta_{\rho}(\bar\varphi_1\star_\theta \bar\varphi_2)}{} = \mu_\theta [({\mathcal F}(\Delta \delta_{\rho}) {\mathcal F}^{-1})((\bar\varphi_1\otimes \bar\varphi_2)))].\label{drintwist}
\end{gather}
Thus, the undeformed coproduct of the symmetry Hopf algebra $U(\hat{\mathcal V})$ is
related to the Drinfeld twist $\Delta^{\mathcal F}$ by the inner endomorphism
$\Delta^{\mathcal F}\delta_\rho := ({\mathcal F}(\Delta \delta_{\rho}) {\mathcal F}^{-1})$
and, by virtue of (\ref{drintwist}), it preserves the covariance:
\begin{gather*}
\delta_\rho \triangleright ((\bar\varphi_1\cdot \bar\varphi_2))) = \mu\circ [\Delta (\delta_\rho)(\bar\varphi_1\otimes \bar\varphi_2))]=
(\delta_{\rho (1)} \triangleright \bar\varphi_1)\cdot (\delta_{\rho (2)} \triangleright \bar\varphi_2)\nonumber\\
  \phantom{\delta_\rho \triangleright ((\bar\varphi_1\cdot \bar\varphi_2)))}{} \stackrel{\theta}{\rightarrow}  \delta^\star_\rho \triangleright (\bar\varphi_1\star_\theta \bar\varphi_2)=
(\delta^\star_{\rho (1)}\: \triangleright \bar\varphi_1)\star_\theta (\delta^\star_{\rho (2)})\: \triangleright \bar\varphi_2),
\end{gather*}
where we have used the Sweedler notation for the coproduct. Consequently, the twisting of the
coproduct is tied to the deformation $\mu\rightarrow \mu_\theta$ of the product when the last
one is def\/ined by
\begin{gather*}
\bar\varphi_1\star_\theta \bar\varphi_2 := ({\mathcal F}^{-1}_{(1)}\:\triangleright \bar\varphi_1) ({\mathcal F}^{-1}_{(2)}\:\triangleright \bar\varphi_2).
\end{gather*}

We want to reiterate at this point that the $\star$-product, associated with the algebra ${\mathcal A}_\theta $,
that we have been considering here is the one originated when considering in turn the f\/lat-Minkowski
space-time quantum mechanics generated by the extended Weyl--Heisenberg group~$H_5$, for the
even more particular case of an extension of the Lie algebra of~$H_5$ by the commutator $[X^\mu , X^\nu]=i\theta^{\mu\nu} $,
for the simplest case when $\theta^{\mu\nu}={\rm const}$. In this case
the generators $\delta_{\rho}$  of isometries become the inf\/initesimal
generators of the Poincar\'e group of transformations, and the coproduct def\/ined in this equation reduces to the
twisted coproduct considered by e.g.~\cite{chaichian4} (see also e.g.~\cite{Chaichian1} and~\cite{chaichian4+,chaichian4++}). Since the embedding coordinates in the canonical
parametrized theory can in general be associated to a curved space-time manifold and, since the constraints and related dif\/feomorphisms
are constructed for such spaces, it seems possible in principle that our formalism could be extended to
curved space-time backgrounds with a $\star$-product determined by the Lie algebra
associated with, for instance, a given homogeneous space.
This would imply f\/inding f\/irst the equivalent of the mapping~(\ref{dar1})
and also, of course, the realization of this map in terms of the $\star$-product, perhaps by a procedure based on the deformation quantization formalism developed by Stratonovich~\cite{Strato}.
A fairly  simple example of the above is the
Darboux map given in \cite{Sanpedro}, for the case of the Snyder algebra~\cite{Snyder}. However, f\/inding a full
realization of the $\star$-product is a~more dif\/f\/icult job.

In equation~(\ref{multpf}) of the previous section we derived the
expression for the inf\/initesimal gauge transformation on a product
of particle f\/ields in  ${\mathcal A}$. Let us now consider the
ef\/fect of such a gauge transformation on the product of two particle
f\/ields in ${\mathcal A}_\theta $ when we have space-time
noncommutativity. For this purpose we f\/irst recall equation~(\ref{partf2})
which shows that if $\varphi$ is a~particle f\/ield, so is its gauge
transformation by pull-back, i.e.\ $\varphi\in
C(P,V)\Rightarrow \varphi' :=f^{\ast}\varphi\in C(P,V)$. From this
it follows that to a given element of $C(P,V)$ we can always
associate another one which is the pull-back of the former, thus the
twisted product of the pull-back with the section $\sigma_u$ of any
pair of particle f\/ields can be written as \begin{gather*}
\bar\varphi'_1 \star_\theta \bar\varphi'_2 =
(\sigma^\ast_u(f^\ast\varphi_1))\star_\theta
(\sigma^\ast_u(f^\ast\varphi_2)). \end{gather*} Observe however that, because
of the noncommutativity that the algebra (\ref{starcomm}) of the embedding
coordinates is required to satisfy, the pull-back to $\mathcal M$
of the gauge transformation (\ref{gauge9}) now should be understood
as $\sigma^\ast_uf^{\ast} \varphi= \bar\zeta^{-1}_{\star}(X)\star_{\theta}\bar\varphi(X)$; so that
\begin{gather}\label{modgt6}
\bar\varphi'_1 \star_\theta \bar\varphi'_2 =(\bar\zeta^{-1}_{\star}\star_{\theta}\bar\varphi_1)
\star_\theta (\bar\zeta^{-1}_{\star}\star_{\theta}\bar\varphi_2),
\end{gather}
where, due to the noncommutativity, equation~(\ref{expmap})
is replaced by
\begin{gather*}
\bar\zeta^{-1}\leadsto\bar\zeta^{-1}_{\star}=\exp_\star(t\bar{\boldsymbol\alpha}(X)):= 1+ t\bar{\boldsymbol\alpha} +
\frac{t^2}{2}\bar{\boldsymbol\alpha}\star_\theta \bar{\boldsymbol\alpha} +\cdots.
\end{gather*}
Using the inf\/initesimal version of this map  we have that $\bar\varphi'_1 =\bar\varphi + \bar{\boldsymbol\alpha}\star_{\theta} \bar\varphi$,
so that (\ref{modgt6}) becomes
 \begin{gather}\label{twistgm1b}
\delta_{\bar\alpha} : (\bar\varphi_1 \star_\theta \bar\varphi_2):=\bar\varphi'_1 \star_\theta \bar\varphi'_2 =
(\bar{\boldsymbol\alpha}(X)\star_{\theta}\bar\varphi_{1}(X))\star_\theta
\bar\varphi_2 +\bar\varphi_1 \star_\theta
(\bar{\boldsymbol\alpha}(X)\star_{\theta} \bar\varphi_{2}(X)).
\end{gather}
By a similar argument, since $f\in GA(P)$ also maps connections into
connections, its inf\/initesimal action on the $\star$-product of two gauge
f\/ields (c.f.\ (\ref{defvwmult})) goes into
\begin{gather*}
\delta_{\bar\alpha}:\left([A, A']_{\star_\theta}({\bf X}_1 ,{\bf X}_2 )\right):=
- \left[\left(d\bar\alpha^A({\bf X}_1) +\frac{1}{2}{c^A}_{CD}[A^C({\bf X}_1), \bar\alpha^D({\bf X}_1)]_{\star_\theta}\right)
\star_{\theta}A'^B({\bf X}_2)\right.\nonumber\\
\left.\qquad{}-\left(d\bar\alpha^A({\bf X}_2) +\frac{1}{2}{c^A}_{CD}[A^C({\bf X}_2), \bar\alpha^D({\bf X}_2)]_{\star_\theta}\right)\star_{\theta}A'^B({\bf X}_1) \right.\nonumber\\
\left.\qquad{}+A^A({\bf X}_1)\star_\theta \left(d\bar\alpha^B({\bf X}_2) +\frac{1}{2}{c^B}_{CD}[A'^C({\bf X}_2), \bar\alpha^D({\bf X}_2)]_{\star_\theta}\right)\right.\nonumber\\
\left.\qquad{}-A^A({\bf X}_2)\star_\theta \left(d\bar\alpha^B({\bf X}_1) +\frac{1}{2}{c^B}_{CD}[A'^C({\bf X}_1), \bar\alpha^D({\bf X}_1)]_{\star_\theta}\right)\right]\otimes [T_A, T_B].
\end{gather*}
Note that we have written the last two equations for the general case of any group of gauge transformations, where $\bar{\boldsymbol\alpha}(X)=\bar\alpha^B T_B$, in order to underline
the fact that, because of the $\star$-product in the multiplication of the f\/ields one needs to apply the constraint that these NC gauge
groups have to be in the fundamental or adjoint unitary representation (i.e.\ $T_A \in U(n)$), since only in this representation the gauge group closes
(c.f.\ e.g.~\cite{Chaichian3, pres}). See however also \cite{arai} for arguments tending to circumvent this constraint).
Hence, in the NC case the generators of gauge symmetry act on particle f\/ields with the fundamental representation
\begin{gather}\label{frep}
\bar\varphi\leadsto \bar\varphi'= \zeta^{-1}_{\star}\star_\theta \bar\varphi=\exp_\star(t\bar{\boldsymbol\alpha}(X))\star_\theta \bar\varphi,
\end{gather}
while on gauge f\/ields the action is via the adjoint representation
\begin{gather}\label{adrep}
A({\bf X}) \leadsto A'({\bf X})=\zeta^{-1}_{\star}\star_\theta A({\bf X}) \star_{\theta} \zeta_{\star} +\zeta^{-1}_{\star}\star_{\theta} (d\zeta_{\star}) ({\bf X}).
\end{gather}
Equations (\ref{frep}) and (\ref{adrep}) agree with those on which \cite{Chaichian2} is based when remarking on
some of the conclusions on deformed gauge theories arrived at in \cite{Wess3, vass, Wess4, giller}.
Indeed, one basic idea in this other approach of gauge twisted theories is the assumption that the gauge generators
$\delta_{\bar\alpha}:=\bar{\boldsymbol\alpha}(X)=\bar\alpha^B(X) T_B$
act on particle and gauge f\/ields with the usual point product, so instead of (\ref{twistgm1b}) they def\/ine
\begin{gather}\label{abelian1}
\delta_{\bar\alpha} (\bar\varphi_1 \star_\theta \bar\varphi_2)
:=(\delta_{\bar\alpha}\bar\varphi_1) \star_\theta \bar\varphi_2 + \bar\varphi_1 \star_\theta (\delta_{\bar\alpha}\bar\varphi_2).
\end{gather}
 Moreover, by assuming that the algebra of the gauge generators can be given an additional Hopf bialgebra structure,
and that the derivatives of any order of the
gauge and particle f\/ields are, as noted in \cite{Chaichian2}, in the same representation of the gauge algebra as
the f\/ields themselves, one could further write
\begin{gather}
\delta_{\bar\alpha} (\bar\varphi_1 \star_\theta \bar\varphi_2)
=(\bar{\boldsymbol\alpha}(X) \bar\varphi_1) \star_\theta \bar\varphi_2 + \bar\varphi_1 \star_\theta \bar{\boldsymbol\alpha}(X)\bar\varphi_2.\nonumber\\
\phantom{\delta_{\bar\alpha} (\bar\varphi_1 \star_\theta \bar\varphi_2)}{}=\mu\circ(\delta_{\bar{\alpha}}\otimes1 +1 \otimes \delta_{\bar{\alpha}})\circ
(e^{\frac{i}{2}\theta^{\mu\nu} \partial_{\mu}\otimes \partial_{\nu}} \bar\varphi_1 \otimes \bar\varphi_2)\nonumber\\
\phantom{\delta_{\bar\alpha} (\bar\varphi_1 \star_\theta \bar\varphi_2)}{} =\mu_\theta [(\Delta^{\mathcal F} \delta_{\bar\alpha}) \circ(\bar\varphi_1\otimes \bar\varphi_2)].\label{abelian1b}
\end{gather}
Assuming a scalar particle f\/ield for simplicity and setting $\bar\varphi_2=\partial_\mu \bar\varphi$ and $\bar\varphi_{1} = \partial_\mu\bar\varphi^\dagger$, it can be readily seen that
one immediate consequence of the extra assumption leading to equating the last two lines in (\ref{abelian1b}) with the f\/irst one is that the latter then yields:
\begin{gather*}
\delta_{\bar\alpha} (\partial_{\mu}\bar\varphi^\dagger\star_\theta \partial_{\mu}\bar\varphi)=0,
\end{gather*}
which implies that the kinetic terms in the Lagrangian of the particle f\/ields are invariant by themselves, so there
would be no need to introduce the gauge potentials to achieve gauge invariance of the theory.
Consequently, since (\ref{abelian1b}) only fully agrees with (\ref{twistgm1b}) when
$\bar{\boldsymbol\alpha}$ is coordinate independent,
there appears to be a discrepancy as a consequence of local internal symmetry between assuming the validity of (\ref{abelian1})
and some essential aspects of the theory of gauge inva\-riance.

Recall furthermore, that a Drinfeld twist (c.f.\ e.g.~\cite{bloh, bloh+,kassel}) involves a~simultaneous and covariant deformation of the product of an algebra ${\mathcal A}$
of functions and the coproduct of a~bialgebra $H$. More specif\/ically, the algebra ${\mathcal A}$ is a
module algebra ($H$-module algebra) over a Hopf bialgebra whose elements are in the universal enveloping algebra $U(L)$
of a Lie algebra~$L$, such that if $x\in L$ then $\Delta(x) = x\otimes 1 + 1\otimes x$, and $x(ab)=x(a)b + ax(b)$ $\forall
\;a,b \in {\mathcal A}$, so that $x$ acts as a~derivation. On the other hand, as shown by equations (\ref{mapgauss}) and (\ref{gauss4}),
the inf\/initesimal gauge transformation of the gauge potential is given by the Poisson bracket of the smeared Gauss constraint $G_\tau[\bar\alpha]$
with the gauge potential; but, as it was also shown in Section~\ref{sec3} of this paper,
the~$\delta_{\bar\alpha}$ can not be made isomorphic to a derivation operator
acting as such on the gauge potentials or particle f\/ields, contrary to the case of the smeared
super-Hamiltonian and super-momenta constraints.
Consequently the algebra of the inf\/initesimal gauge transformations can not be considered as part of
the Hopf algebra of the space-time dif\/feomorf\/isms $\delta_\xi$, associated with Lie algebra $L$ and
its universal envelope, from which a Drinfeld twist could be properly constructed.
Note also that in the context of the canonical parametrized
formalism, the Gauss constraint is def\/ined on the spacelike hypersurface $\Sigma $ and, again contrary to the super-Hamiltonian and super-momenta constraints, does not
depend on the embedding variables. This translates in the fact that for the NC case the space-time dif\/feomorphisms $\delta_\xi$, on the one hand,
and the inf\/initesimal gauge transformations $\delta_{\bar\alpha}$, on the other, act quite dif\/ferently on the gauge and particle f\/ields.
 This is clearly seen when comparing the actions  (\ref{moddif}) and (\ref{frep}) on the gauge and particle f\/ields, as well as their actions
(\ref{drintwist}) and (\ref{twistgm1b}) on their respective products.

It thus appears from our present results as well as from those in \cite{nos1} (where the noncommutative reparametrized scalar f\/ield was considered
and its respective constraints together with their anti-homomorphic relation to space-time dif\/feomorphisms was explicitly established), that
it might not be possible to extend the concept of a Drinfeld twist symmetry to include gauge symmetries,
 when considering the minimal coupling of gauge and particle f\/ields in order to investigate a full model of NC theory in the context
of the canonical reparametrized theory (see e.g.~\cite{Chaichian3} regarding this point). 

 However, if one were to consider relaxing the concept of twisted symmetries and
modify the def\/inition of a deformed Leibniz rule (such as the one exhibited in (\ref{abelian1})), several dif\/ferent twists and gauge
invariants may be constructed that would lead to alternate formulations
for NC gauge theories. Some new ideas in this context that might help to remove some of the inconsistencies pointed out here as well as elsewhere,
are discussed in~\cite{Vazquez,Vazquez+}. This would involve, essentially, assuming dif\/ferent deformations of products of elements in the same algebra of space-time
functions~$\mathcal A$, when considering dif\/ferent transformation groups. Such an assumption however, would be hard to reconcile with the point of view that
the product in this algebra of functions is inherited
from the deformation of the algebra of space-time coordinates and its dynamical origin in the quantum mechanical mini-superspace.


As it was remarked previously the $\star$-product considered so far
applies to an underlying f\/lat Minkowski space-time, and the corresponding twisted isometries
refer then to the Poincar\'e group. It is interesting to observe, however, that our
formalism admits a natural extension of (\ref{starprod}) which allows us to consider
much more general symplectic structures than (\ref{seccom})  that would imply
noncommutativity among all the symplectic variables $z^A=(X^\alpha, A_a, \psi; P_\alpha, \pi^a, \pi)$.
Moreover, because of the appearance of the embedding metric in the canonical parametrized formalism,
this could lead in turn to the possibility of extending our analysis to the case of twisted isometries
on curved space backgrounds.

Even within the f\/lat Minkowski space-time case, we could have a more general
symplectic structure that would lead to a dif\/ferent $\star$-product with bi-dif\/ferentials
involving some of the other f\/ields in the theory.
Consider for instance the symplectic structure resulting in the Dirac brackets:
\begin{gather}
  \{X^\alpha, X^\beta \}^\ast  =  i\theta^{\alpha\beta} ,\qquad
  \{X^\alpha, P_\beta \}^\ast  =  i \delta_\alpha^\beta ,\qquad
  \{P_\alpha, P_\beta \}^\ast  = 0,\nonumber\\
  \{A_a , A_b \}^\ast = 0,  \qquad
  \{A_a , \pi^b \}^\ast  =  i \delta_a^b , \qquad
  \{\pi^a ,\pi^b \}^\ast  =  i\beta^{ab},\label{Hei1}
\end{gather}
(and the remainder equal to zero). Here the Darboux
map, that takes us from the extended algebra
(\ref{Hei1}) to the usual Heisenberg algebra, is given by the transformations:
\begin{gather}\label{Dar1}
  \tilde X^\alpha = X^\alpha + \frac{\theta^{\alpha\beta}}{2} P_\beta,  \qquad
    \tilde \pi^a  = \pi^a + \frac{\beta^{ab}}{2} A_b.
\end{gather}
These maps are unique up to a canonical transformation on the
phase-space $(X^\alpha, P_\alpha, A_a, \pi^a)$. In order to
construct the deformed constraints, note that in the expressions for
$\Phi_0$ and $\Phi_a$ in~(\ref{Cons1}) there appear the projectors
$n^\alpha(\sigma, X)$ and $X^\alpha_a(\sigma, X)$ as well as the
3-metric $\gamma_{ab}$, all of which are functionals of the space-time
embedding coordinates $X^\alpha$. These quantities thus need to be
modif\/ied according to (\ref{Dar1}). On the other hand, the Gauss
constraint also requires to be modif\/ied in order that the Dirac
bracket algebra of the new constraints be the same as the Poisson
algebra of the original ones.
The resulting deformed constraints are then:
\begin{gather*}
\tilde \Phi_0 = P_\alpha \tilde n^\alpha + \frac{1}{2} \tilde
\gamma^{-1/2} \tilde \gamma_{ab}\tilde\pi^a \tilde\pi^b +
\frac{1}{4}
\tilde \gamma^{1/2}\tilde \gamma^{ac}\tilde \gamma^{bd}F_{ab} F_{cd},\nonumber\\
  \tilde \Phi_a      = P_\alpha \tilde X^\alpha_{\ ,a} -F_{ab}\tilde\pi^b,\qquad
  \tilde G = \tilde\pi^a_{\ ,a},
\end{gather*}
where the tilde on top of a symbol denotes the replacement of the
space-time coordinates according to~(\ref{Dar1}). However, one point to
observe is even that the constraints have been
deformed by the fact that their algebra involves now Dirac brackets
instead of Poisson brackets, the Darboux transformations
(\ref{Dar1}) preserve the functional form of their algebra, so they
can still be made anti-homomorphic to an algebra of deformed
space-time dif\/feomorphisms, by a procedure analogous to the one
described here. Also note, in particular, that the original f\/ields
$z^A=(X^\alpha, A_a, \psi; P_\alpha, \pi^a, \pi)$ will now transform
according to the twisted dif\/feomorphisms of the theory. Thus, while
the electric f\/ield $\pi^a$ will no longer be gauge invariant, the
new f\/ield $\tilde \pi^a$ will be, under the gauge transformation
associated with the modif\/ied Gauss constraint. Note also that the
last equation in (\ref{Hei1})implies that the Drinfeld deformation
of the algebra of functions of the f\/ields involves a
$\star$-product which is a composition of (\ref{starprod}) with
 \begin{gather*}
\star_{\beta}:= \exp\left[\frac{i}{2}\beta^{ab}\int
d\sigma''\frac{\overleftarrow\delta} {\delta
\pi^{a}(\tau,\boldsymbol\sigma'')}\frac{\overrightarrow\delta}{\delta
\pi^{b}(\tau,\boldsymbol\sigma'')}\right].
\end{gather*}

\subsection*{Acknowledgements}

The authors are grateful to Prof. Karel Kucha\v{r} for fruitful
discussions and clarif\/ications concerning his work on parametrized
canonical quantization. They are also grateful to the referees for
some very pertinent comments and suggestions which helped to clarify
considerably some points in the manuscript. The authors also acknowledge partial support from
CONACyT projects UA7899-F (M.R.) and  47211-F (J.D.V.) and DGAPA-UNAM grant IN109107 (J.D.V.).

\pdfbookmark[1]{References}{ref}
\LastPageEnding

\end{document}